\documentclass[11pt]{article}
\usepackage{amsmath,amsfonts,amssymb}
\usepackage{latexsym}
\usepackage{dcolumn}
\usepackage{graphicx,epsfig}
\usepackage{amsthm}
\usepackage{color}
\begin{document}
	\setcounter{page}{1}
	
	\pagestyle{plain} \vspace{1cm}
	\begin{center}
\Large{\bf Reheating and particle creation in unimodular $f(R,T)$ gravity}\\
\small \vspace{1cm}{\bf Fateme Rajabi\footnote{fa.rajabi@stu.umz.ac.ir}}\quad and\quad {\bf Kourosh Nozari\footnote{knozari@umz.ac.ir(Corresponding Author)}}\quad\\\vspace{0.25cm}
Department of Theoretical Physics, Faculty of Basic Sciences,\\
University of Mazandaran,\\
P. O. Box 47416-95447, Babolsar, IRAN
	\end{center}
	
	\vspace{1cm}
\begin{abstract}
We study cosmological inflation and reheating in the unimodular $f(R,T)$ gravity. During the reheating era, which takes place just after the end of inflation, the energy density of inflaton is converted to radiation energy through, for instance, rapid oscillation about the minimum of the potential. We quantify our investigation by calculating the reheating temperature. This quantity is written in terms of the spectral index and the power spectrum, which provides a suitable framework to constrain the parameter space of the model. We discuss the massless particle creation for a spatially flat, homogeneous and isotropic universe in the context of unimodular $f(R,T)$ gravity. We obtain the number of created particles per unit volume of space. In order to avoid the complexity of field equations, we conformally transform in Einstein frame and investigate the reheating by considering some specific illustrative examples. Also we obtain the corresponding  analytical solutions in addition to some numerical estimations.
	
\end{abstract}

	\newpage
	
	\section{Introduction}

The idea of cosmological inflation was proposed initially to address at least some of the problems of the standard cosmological model such as the flatness, horizon, magnetic monopoles, large scale isotropy and homogeneity problems~\cite{Gu81, Lin90, Lid96, Bas18, Her17, Yan16, Mar14}. One of the most important predictions of the cosmic inflation is creation of small density inhomogeneity from quantum fluctuations in the early universe ~\cite{Mar92}. In this respect, physics of the early universe can be explained satisfactorily by cosmological inflation. One of the model to explain this era, is to consider a slowly rolling scalar field, named inflaton, whose energy density was dominated by its potential during cosmological inflation ~\cite{Lin83,Kal09,Cre14}.

The reheating stage at the end of inflation is an important part of the cold inflation paradigm. At this era, the inflaton field decayed and converted into the standard model particles. In a simple inflation scenario, the scalar field responsible for inflation rolls slowly on a nearly flat potential and then universe expands quasi-exponentially. At the end of inflation, the slow-roll conditions break down and so the inflaton field oscillates around the minimum of its potential. Eventually, the inflaton field loses its energy by oscillation and transfer this energy to relativistic particles corresponding to radiation dominated universe with ordinary matter~\cite{Abb82, Dol82, Alb82, All10}. The processes of reheating, creation of ordinary matter and transition to the radiation-dominated universe occur by the production of gravitational particles~\cite{Arm99, Kob10, Pee99, Kun12}. Without reheating, the universe would be empty and become cold after the inflation in cold inflation scenario. Several models of perturbative and non-perturbative processes of reheating have been studied by authors which include the perturbative decay of an oscillating inflaton field at the end of inflation~\cite{Abb82, Dol82, Alb82}, the tachyonic instability ~\cite{Gre97, Shu06, Duf06, Abo10, Fel01, Kof01}, parametric resonance decay ~\cite{Kof94, Tre90, Kof97} and the instant preheating ~\cite{Fel99}.

One can characterize the reheating era by studying the number of e-folds ($N_{reh}$) or reheating temperature ($T_{reh}$) and the effective equation of state parameter ($\omega_{eff}$). These quantities give some more constraints on the parameter space of the model ~\cite{Dai14, Mun15, Coo15, Cai15, Uen16, Noz17}. Rapid interactions between inflaton field and other matter in thermal equilibrium can be characterized by reheating temperature, $T_{reh}$. However, the exact value of reheating temperature has not been determined yet, though it has been constrained in both sides in literature. Big bang nucleosynthesis (BBN), large scale structure, light elements abundance and cosmic microwave background (CMB) put the lower bound on reheating temperature, $4$\,MeV $\lesssim T_{reh}$ ~\cite{Han04}. The energy scale at the end of inflation can be considered as the upper bound of reheating temperature which is around the GUT scale, $T_{reh} \lesssim 10^{16}$ GeV. In ~\cite{Mar10} the authors constrained the reheating phase indirectly by involving supersymmetry and considering the gravitino production, and also by the CMB observational data as $6$\,TeV $\lesssim T_{reh} \lesssim 10^{4}$\, TeV. Therefore, $T_{reh}$ is an important parameter that gives some critical information about reheating era.

Another quantity associated with reheating era is the effective equation of state parameter ($\omega_{eff}$) during various stages of reheating that lies between the potential dominance ($\omega_{eff}=-1$ ) and kinetic dominance ( $\omega_{eff}=1$) era ~\cite{Ami15}. It is obvious that the effective equation of state parameter at the end of inflation is $-\frac{1}{3}$ and its value at the beginning of the radiation dominated era is $+\frac{1}{3}$. So we consider the effective equation of state parameter during reheating epoch in the range of $-\frac{1}{3}\leqslant \omega_{eff} \leqslant \frac{1}{3}$. During initial phase of reheating, the oscillation frequency of the massive inflaton will be larger than the expansion rate. Therefore, this leads to vanishing of the effective equation of state parameter of inflaton field in the beginning of the reheating that corresponds to equation of state parameter of the pressureless (dust) matter. So, by oscillation of the inflaton field and converting its energy into other particles, the effective equation of state parameter increases from $0$ to $\frac{1}{3}$ at the beginning of the radiation dominance. After the accelerating expansion, the gravitational particles decay into the plasma of the relativistic particles corresponding to the radiation dominated universe ~\cite{Zel77, Sta81, Vil85, Mij86, For87, Mot12}. In the process of reheating, the inflaton field converts its energy to the ordinary matter through coupling to these fields. However, to avoid large coupling during the reheating epoch, there is significant fine-tuning problem of couplings ~\cite{Oik17}. In this perspective, modified gravity is an alternative theory to describe reheating phase. In this theory, gravity acts as the effective equation of state of the matter fields which produces non-trivial effects on the field equations and leads to the heating of the matter content of the universe ~\cite{Mot12, Oik17, Goo18, Mie11}.

Observational data, such as the Supernova Ia (SN Ia) and Cosmic Microwave Background (CMB) data all obviously indicated that the universe currently is in a phase of positively accelerated expansion ~\cite{Per99, Rie98, Rie99, Han00, Pee03}. General Relativity cannot describe the early and late time acceleration of the universe by just ordinary matter fields. So, some kinds of modifications should be imposed on general relativity to address the large scale cosmic speed up. General relativity does not work as a fundamental theory in the framework of Quantum Field Theory (QFT); this theory cannot be quantized. To address these fundamental problems, several theories have been proposed ~\cite{Fra85, Wit86, Fri86, Bar04, Bri04}. However, the quantum effects can be considered in gravitational theories. Those effects can be attributed to the terms of trace of the energy-momentum tensor in the gravitational part of the $f(R)$ action which is named $f(R,T)$ theories ~\cite{Mor16}, that firstly was proposed in Ref.~\cite{Har11}. This theory is studied extensively in several literature ~\cite{Hou12, Pia12, Mom12, Noz18, Alv13, Zub15, Nou15, Alv16, Mor15, Sin14, Far15, Sah14, Red14, Sah16, Mom16, Bha20}. The dependence on $T$ can be induced by exotic fluid or quantum effects. In this theory, the equations of motion show the presence of an extra force acting on the test particles, and the motion are generally non-geodesic. Also, the non-minimal coupling of the Ricci scalar and matter Lagrangian density have been studied in ~\cite{Ber07, Har08, Tha11, Ber08, Boe08, Far09, Har10}.

 On the other hand, an alternative theory to address the cosmological constant problem is the \emph{unimodular gravity} which was firstly proposed by Einstein~\cite{Wei89, Ein19}. In this scenario, the cosmological constant arises from the trace-free part of the Einstein's field equations. In this theory, the determinant of the metric has no dynamics and thus the cosmological constant originates naturally as an integration constant. So, the cosmological constant is not present in the field equations. Another interesting feature of unimodular gravity is that the problem of late-time accelerated expansion can be addressed naturally ~\cite{Jai12, Kar12}. Also, the unimodular gravity is capable to explain the inflation era ~\cite{Cho15, Bam16}. The extension of unimodular gravity to modified theories of gravity such as $f(R)$ gravity, $f(T)$ (modified teleparallel gravity with $T$ as torsion scalar) and $f(R,T)$ (with $T$ as the trace of the energy-momentum tensor) have been studied in Refs.~\cite{Noj16, Odi16, Noji16, Odin16, Sae16, Nas16, Bam16, Raj17}. At the classical level, the unimodular gravity is equivalent to general relativity, but may provide differences when
quantum considerations are taken into account~\cite{Alv05, Smo11, Smo09, Unr89, Kuc91, Alv15, Fio10, Eic13, Sal14, Alv07}. Some other aspects of unimodular gravity have been addressed in Refs.~\cite{Zee85, Hen89, Ng91, Fin01, Uza11, Bar14, Bur15, Sha09, Boc07}.

In this paper, we investigate cosmic inflation and reheating stage in the context of our proposed unimodular $f(R,T)$ gravity~\cite{Raj17} (see also~\cite{Raj21} for the energy conditions in this modified gravity scenario). By adopting the Jordan frame, we study the slow roll inflation, the rapid oscillation phase, the radiation dominated and the recombination eras and calculate the e-folds numbers during these epochs. Then, we compute the reheating temperature as a function of the spectral index and the power spectrum. We study also the particle production in this model for a flat, homogeneous and isotropic universe. To be more clarified, in each step we provide some numerical estimations of the main related physical quantities. Finally, to avoid complexity of the field equations and in analogy to scalar fields model, we work in the Einstein frame. Finally and as a clarifying example, we investigate reheating process for $f(R,T)=R+\alpha R^2+\gamma T$ model in Einstein frame and obtain its reheating temperature and other important quantities with some numerical estimations.

The structure of this paper is as follows: In section 2, we briefly review our previously proposed unimodular $f(R,T)$ gravity~\cite{Raj17}, and provide some essential information of this model in inflationary era. In section 3 we analyze the inflation and reheating for power law potential by studying the
evolution of the universe. We derive some expressions for e-folds number and temperature in this era in terms of the scalar spectral index and power spectrum. In section 4, particle production phenomenon for massless particles in a radiation dominated universe in the framework of unimodular $f(R,T)$ theories of gravity is investigated. In section 5 we obtain the analytical solutions of the inflation and reheating in the Einstein frame. Section 6 is devoted to summary and conclusions. \\

\section{The unimodular $f(R,T)$ gravity and inflation}
In this section we firstly review briefly the main ideas behind the unimodular $f(R,T)$ gravity and then we reveal some essential features of this model in inflationary era. The original motivation for introducing unimodular gravity was to solve the cosmological constant problem. The basic idea of unimodular gravity is that the determinant
of the metric must follow a specific constraint. In fact, the spacetime metric determinant is not dynamical and is fixed as
\begin{equation}
\sqrt{-g}=1\,.
\end{equation}
We consider a spatially flat Friedmann-Robertson-Walker (FRW) space-time with the following metric
\begin{equation}
ds^{2}= -dt^{2} + a^{2}(t)dx_{i}dx^{i},\quad i=1,2,3\,,
\end{equation}
where $a(t)$ is the scale factor. This metric does not satisfy the unimodular condition (1). So, following Ref.~\cite{Noj16}, we redefine the cosmological time $t$ as
\begin{equation}
d\tau=a^3(t)dt\,.
\end{equation}
In this case, we can rewrite  the metric (2) as follows
\begin{equation}
ds^{2}= -a^{-6}(\tau)d\tau^{2} +a^2(\tau)dx_{i}dx^{i},\quad i=1,2,3\,.
\end{equation}
In Ref.~\cite{Raj17} we constructed an extension of unimodular gravity named unimodular $f(R,T)$ gravity, where $R$ and $T$ are the Ricci scalar and the trace of the energy-momentum tensor, $T_{\mu\nu}$, respectively. Also, the cosmic inflation and cosmological reconstruction of the unimodular $f(R,T)$ gravity were discussed there. The unimodular action of the model in the Jordan frame gets the following form
\begin{equation}
S=\frac{1}{2\kappa^2}{\int}{d^4x\Big[\sqrt{-g}f(R,T)-2\lambda(\sqrt{-g}-1)\Big]}
+{\int}d^4x\sqrt{-g}{\cal{L}}_{m}\,\,,
\end{equation}
where ${\cal{L}}_{m}$ describes the matter Lagrangian density and depends only on the matter fields and metric tensor components $g_{\mu\nu}$. We imposed the unimodular constraint by inserting the Lagrange multiplier, $\lambda$, in the action. Hence, we can obtain unimodular constraint (1) by varying the action (5) with respect to the Lagrange multiplier. The energy-momentum tensor of matter is
\begin{equation}
T_{\mu\nu}=-\frac{2}{\sqrt{-g}}\frac{\delta(\sqrt{-g}{\cal{L}}_{m})}{\delta
g^{\mu\nu}}=g_{\mu\nu}{\cal{L}}_{m}-2\frac{\partial{\cal{L}}_{m}}{\partial g^{\mu\nu}}\,.
\end{equation}
We find the field equations by varying the action (5) with respect to the metric
\begin{eqnarray}
f_{,R}(R,T)R_{\mu\nu}-\frac{1}{2}g_{\mu\nu}f(R,T)+(g_{\mu\nu}\Box-\nabla_{\mu}\nabla_{\nu})f_{,R}(R,T)+\lambda g_{\mu\nu}=\\ \nonumber
\kappa^2 T_{\mu\nu}
-f_{,T}(R,T)(T_{\mu\nu}+\Theta_{\mu\nu})\,,
\end{eqnarray}
where $f_{,R}(R,T)$ and $f_{,T}(R,T)$ denote partial derivatives of $f(R, T)$ with respect to $R$ and $T$ respectively. $\Theta_{\mu\nu}$ is defined as
\begin{equation}
\Theta_{\mu\nu}\equiv g^{\alpha\beta} \frac{\delta
T_{\alpha\beta}}{\delta g^{\mu\nu}}=-2T_{\mu\nu}+g_{\mu\nu}{\cal{L}}_{m}
-2g^{\alpha\beta}\frac{{\partial^2}{\cal{L}}_{m}}{\partial g^{\mu\nu}\partial g^{\alpha\beta}}\,.
\end{equation}
By taking the trace of the field equation (7) we get
\begin{equation}
f_{,R}(R,T)R+3\Box f_{,R}(R,T)-2f(R,T)+4\lambda=\kappa^2
T-f_{,T}(R,T)(T+\Theta)\,.
\end{equation}
Then, the field equation (7) leads to
\begin{eqnarray}
&f_{,R}(R,T)\Big[R_{\mu\nu}-\frac{1}{3}g_{\mu\nu}R\Big]
-\nabla_{\mu}\nabla_{\nu}f_{,R}(R,T)+\frac{1}{6}
f(R,T)g_{\mu\nu}-\frac{1}{3}\lambda
g_{\mu\nu}=\nonumber\\
&\kappa^2\Big[T_{\mu\nu}-\frac{1}{3}g_{\mu\nu}T\Big]
-f_{,T}(R,T)\Big[T_{\mu\nu}-\frac{1}{3}g_{\mu\nu}T\Big]
-f_{,T}(R,T)\Big[\Theta_{\mu\nu}-\frac{1}{3}g_{\mu\nu}\Theta\Big]\,.
\end{eqnarray}
This equation shows that the unimodular $f(R,T)$ gravity corresponds to normal $f(R,T)$ equations with an additional
cosmological constant.

The covariant derivative of the
field equations (7) gives
\begin{eqnarray}
\nabla^{\mu}T_{\mu\nu}=\frac{1}{\kappa^2-f_{,T}(R,T)}
\Big[-\frac{1}{2}g_{\mu\nu}f_{,T}(R,T)\nabla^{\mu}T+(T_{\mu\nu}
+\Theta_{\mu\nu})\nabla^{\mu}f_{,T}(R,T)\\ \nonumber+f_{,T}(R,T)\nabla^{\mu}\Theta_{\mu\nu}
+g_{\mu\nu}\nabla^{\mu}\lambda\Big]\,,
\end{eqnarray}
where we have used the relation
$(\nabla_{\mu}\Box-\Box\nabla_{\mu})f_{,R}=R_{\mu\nu}\nabla^{\nu}f_{,R}$.
The above equation implies that the stress-energy tensor of the matter fields is not conserved. This is due to the exchange of energy and momentum between geometry and matter, as a consequence of the geometry-matter coupling encapsulated in the trace, $T$, of the energy-momentum tensor of the matter field. Also, equation (11) expresses that the form of the matter Lagrangian affects not only the energy-momentum conservation, but also the particle motion, see \cite{Raj17} for more details.

We consider a perfect fluid as the source of the energy-momentum tensor with
\begin{equation}
T_{\mu\nu}=(\rho_{m}+p_{m})u_{\mu}u_{\nu}+p_{m} g_{\mu\nu}\,,
\end{equation}
where $u_{\mu}$ is the four-velocity of perfect fluid, $u_{\mu}u^{\mu}=-1$. $\rho_{m}$ and $p_{m}$ are respectively the energy density and pressure of the perfect fluid with equation of state $p_{m}={\omega}\rho_{m}$. Therefore, by comparing the equations (6) and (12), the matter Lagrangian can be set as
${\cal{L}}_{m}=p_{m}$ ~\cite{Har11}. So $\Theta_{\mu\nu}$ takes the following form
\begin{equation}
\Theta_{\mu\nu}=-2T_{\mu\nu}+p_m g_{\mu\nu}\,,
\end{equation}
Substituting this equation into the field equation (7) we have
\begin{eqnarray}
f_{,R}(R,T)R_{\mu\nu}-\frac{1}{2}g_{\mu\nu}f(R,T)+\big(g_{\mu\nu}\Box-\nabla_{\mu}\nabla_{\nu}\big)f_{,R}(R,T)
+\lambda g_{\mu\nu}=\kappa^2 T_{\mu\nu}+f_{,T}(R,T) T_{\mu\nu} \\ \nonumber
-p_m g_{\mu\nu}f_{,T}(R,T)\,,
\end{eqnarray}
For the spatially flat FRW metric (4), the Friedmann equations are given as
\begin{eqnarray}
-(3\dot{{\cal{H}}}+12{\cal{H}}^2)f_{,R}+\frac{1}{2}(f-2\lambda)a^{-6}+3{\cal{H}}\dot{f}_{,R}=
[\kappa^2\rho_m+f_{,T}(\rho_m+p_m)]a^{-6}
\end{eqnarray}
\begin{eqnarray}
(\dot{{\cal{H}}}+6{\cal{H}}^2)f_{,R}-\frac{1}{2}(f-2\lambda)a^{-6}-5{\cal{H}}\dot{f}_{,R}
-\ddot{f}_{,R}=\kappa^2p_ma^{-6}\,,
\end{eqnarray}
where ${{\cal{H}}}$ is the modified Hubble parameter and a dot denotes derivative with respect to $\tau$. By using the unimodular FRW metric (4), equation (11) takes the following form	
\begin{eqnarray}
\dot{\rho}_m+3{\cal{H}}(\rho_m+p_m)=-\frac{1}{\kappa^2
+f_{,T}}\Big[-\frac{1}{2}\dot{T}f_{,T}+(\rho_m+p_m)
\dot{f}_{,T}+\dot{p}_{m}f_{,T}+\dot{\lambda}\Big]\,.
\end{eqnarray}
This equation shows that a general unimodular $f(R,T)$ model does not satisfy the usual conservation law. as we will see, this non-conservation affects the rate of gravitational particle creation.

Now, we are going to analyze the behavior of unimodular $f(R,T)$ theory in the presence of inflaton field. So, we consider the Lagrangian density of matter as
\begin{equation}
{\cal{L}}_m=-{\frac{1}{2}}{\partial_{\mu}\phi}\,{\partial^{\nu}\phi}-V(\phi)\,,
\end{equation}
where $\phi\equiv \phi(\tau)$. By using the equation (12), the density and pressure of the inflaton field are given as
\begin{equation}
\rho=\frac{1}{2}\dot{\phi}^2{a^6}+V(\phi)\,,
\end{equation}
\begin{equation}
p=\frac{1}{2}\dot{\phi}^2{a^6}-V(\phi)\,,
\end{equation}
 respectively. We consider the simplest $f(R,T)$ gravity as $f(R,T)=R+\beta T$ with constant $\beta$, where its viability has been studied in Refs.~\cite{Har11, Mor19, Mor17, Azi13}. Also, following our previous study, Ref.~\cite{Raj17}, we assume the Lagrangian multiplier, $\lambda$, to be as follows
\begin{equation}
\lambda={\alpha_1}R+{\alpha_2}T\,,
\end{equation}
where $\alpha_1$ and $\alpha_2$ are constant. So, the modified Friedman equations now are given as
\begin{eqnarray}
-6{\alpha_1}\dot{H}-(12\alpha_1-3)H^2=\frac{1}{2}(\kappa^2+\beta+2\alpha_2){\dot{\phi}}^2+(\kappa^2+2\beta-4\alpha_2)V \,,
\end{eqnarray}
\begin{eqnarray}
(6\alpha_1-2)\dot{H}-(3-12\alpha_1)H^2=\frac{1}{2}(\kappa^2+\beta-2\alpha_2){\dot{\phi}}^2-(\kappa^2+2\beta-4\alpha_2)V\,,
\end{eqnarray}
where $\dot{H}\equiv\frac{dH}{dt}$. Note that in the above equations we have used $d\tau=a^3(t)dt$. Moreover, the scalar field obeys the equation of motion as
\begin{equation}
\big[\kappa^2+\beta-2\alpha_2-6\alpha_1(\beta+\kappa^2)\big]\ddot{\phi}+(3-12\alpha_1)(\beta+\kappa^2)H\dot{\phi}+(\kappa^2+2\beta-4\alpha_2)V^{\prime}(\phi)=0\,,
\end{equation}
where a prime remarks derivative with respect to $\phi$. During the inflation phase, the slow roll conditions which are characterized by $\frac{\dot{H}}{H^2}\ll1$ and $\ddot{\phi}\ll 3H\dot{\phi}$, imply that
\begin{eqnarray}
H^2=\frac{\kappa^2+2\beta-4\alpha_2}{3(1-4\alpha_1)}V(\phi)\,,\qquad\quad\,\nonumber\\
H\dot{\phi}=-\frac{\kappa^2+2\beta-4\alpha_2}{3(1-4\alpha_1)(\beta+\kappa^2)}V^{\prime}(\phi)\,\,.
\end{eqnarray}
For the power law  potential $V(\phi)=q \,{\phi}^n$, the above equations become
\begin{equation}
H=\sqrt{\frac{\kappa^2+2\beta-4\alpha_2}{3(1-4\alpha_1)}q\phi^n}\,\,,\qquad\quad
\end{equation}
\begin{equation}
H\dot{\phi}=-\frac{\kappa^2+2\beta-4\alpha_2}{3(1-4\alpha_1)(\beta+\kappa^2)}nq\phi^{n-1}\,\,.
\end{equation}
By substituting equation (26) into equation (27) and then by integrating the result, we obtain
\begin{equation}
\phi=\Bigg[\phi_i^{\frac{4-n}{2}}+\frac{n(n-4)}{2(\beta+\kappa^2)}\sqrt{\frac{q(\kappa^2+2\beta-4\alpha_2)}{3(1-4\alpha_1)}}t\Bigg]^{\frac{2}{4-n}}\,.
\end{equation}
Consequently, the Hubble parameter becomes as
\begin{equation}
H=\sqrt{\frac{q(\kappa^2+2\beta-4\alpha_2)}{3(1-4\alpha_1)}}\Bigg[\phi_i^{\frac{4-n}{2}}+\frac{n(n-4)}{2(\beta+\kappa^2)}
\sqrt{\frac{q(\kappa^2+2\beta-4\alpha_2)}{3(1-4\alpha_1)}}t\Bigg]^{\frac{n}{4-n}}\,.
\end{equation}
We note that the limit $\beta=\alpha_1=\alpha_2=0$ retrieves the standard general relativistic results and $\alpha_1=\alpha_2=0$ recovers the $f(R,T)$ gravity results~\cite{Bha20}. Moreover, we can determine the scale factor as follows
\begin{equation}
a=a_{i}\exp\Bigg\{-\Big(\frac{\beta+\kappa^2}{2n}\Big)\bigg[\phi_i^{\frac{4-n}{2}}+\frac{n(n-4)}{2(\beta+\kappa^2)}
\sqrt{\frac{q(\kappa^2+2\beta-4\alpha_2)}{3(1-4\alpha_1)}}t\bigg]^{\frac{4}{4-n}}\Bigg\}\,,
\end{equation}
that is a constant at $t=0$. If $\phi_i=0$, then equation (30) reduces to the well-known expression for the intermediate scale factor in the form of $a(t)=\exp\,(Bt^f)$ ~\cite{Bar19,Nun20}. Figure 1 shows the evolution of the scale factor as a function of time with $f(R,T)=R+\beta T$ and Lagrange multiplier as $\lambda=\alpha_1R+\alpha_2T$, for three different values of the parameters $n$, $\beta$ and $\alpha_1$. This figure shows qualitatively the possibility of having positively accelerated expansion of the universe in this model.
\begin{figure*}
\begin{center}
{\includegraphics[width=0.53\textwidth,origin=c,angle=0] {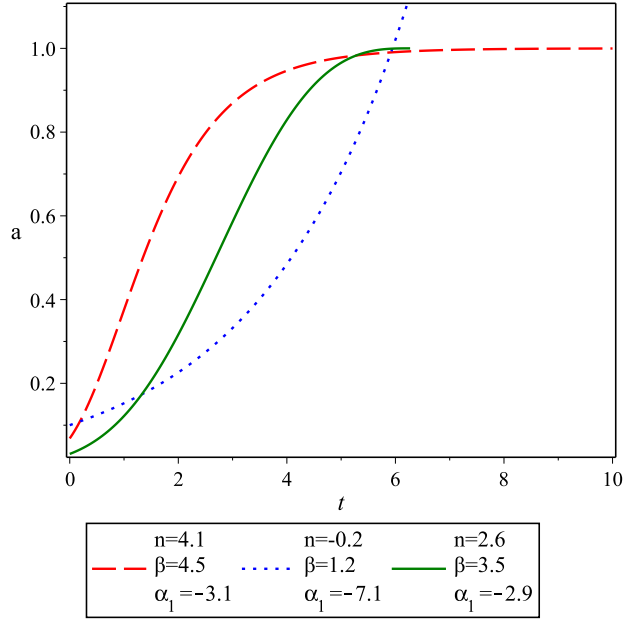}}
\end{center}
\caption{\small{Typical evolution of the scale factor versus the cosmic time in $f(R,T)=R+\beta T$ with potential $V(\phi)=q\phi^n$ and for some selected values of $n$, $\beta$ and $\alpha_1$. To plot this figure we have set $\alpha_{2}=-1.9$ and $q=1$. The blue curve (dotted-line) shows the intermediate scale factor for $\phi_i=0$.}}
\end{figure*}
In order to study inflation era, it is useful to define the slow-roll parameters as
\begin{equation}
\varepsilon_1\equiv -\frac{\dot{H}}{H^2}=\frac{1}{2(\beta+\kappa^2)}\bigg(\frac{V^{\prime}(\phi)}{V(\phi)}\bigg)^2\,,\hspace{1.7cm}
\end{equation}
\begin{equation}
\varepsilon_2\equiv-\frac{2\dot{H}}{H^2}+\frac{\ddot{H}}{H\dot{H}}= 4\varepsilon_1 \Bigg(1-\frac{V^{\prime\prime}(\phi)V(\phi)}{V^{\prime2}(\phi)}\Bigg)\,.
\end{equation}
The number of e-folds during the cosmic evolution from the beginning of inflation, $t_i$, until the end of inflation, $t$, is given by
\begin{equation}
N=\int _{t_i}^ {t}H dt= \int _{\phi_i}^ {\phi} \frac{H}{\dot{\phi}}d\phi= -(\beta+\kappa^2)\int _{\phi_i}^ {\phi} \frac{V(\phi)}{V^{\prime}(\phi)}d\phi\,,
\end{equation}
where the relations in equation (25) are used. For the power law potential, the last expression takes the following form
\begin{equation}
N=\frac{(\beta+\kappa^2)}{2n}(\phi_i^2-\phi^2)\,.
\end{equation}
Now by setting $\varepsilon_{1}=1$ at the end of inflation, we get $\phi_{e}=\frac{n}{\sqrt{2(\beta+\kappa^{2})}}$. So, we find
\begin{equation}
\phi_{i}=\sqrt{\frac{n(4N+n)}{2(\beta+\kappa^2)}}\,.
\end{equation}
These equations will be very useful to obtain the observable inflationary parameters. With these relations, equations (31) and (32) reduce to the following forms
\begin{equation}
\varepsilon_1=\frac{n^2}{2(\beta+\kappa^2){\phi}_{i}^2}=\frac{n}{4N+n}\,,
\end{equation}
\begin{equation}
\varepsilon_2= \frac{4}{n}\varepsilon_1 =\frac{4}{4N+n}\,,\qquad\,\,
\end{equation}
Also, we obtain the observable inflationary parameters as
\begin{equation}
r=16\varepsilon_1=\frac{16n}{4N+n}\,,\hspace{2.4cm}
\end{equation}
\begin{equation}
n_s=1-2(\varepsilon_1+\varepsilon_2)=1-\frac{2(n+4)}{4N+n}\,,
\end{equation}
\begin{equation}
n_T=-2\varepsilon_1=-\frac{2n}{4N+n}\,.\hspace{2cm}
\end{equation}
With these equations, we are in the position to confront this model with observational data. For instance, taking $n=0.8$ and for e-folding $N=60$, we find $r\approx0.07$, $n_s\approx0.96$, and $n_T\approx-0.009$. These numerical results for the inflationary observables show that this model, for some specific and viable values of the model's parameters, can be in good agreement with observational data from Planck2018 probe~\cite{Planck2018}.

To calculate the perturbation parameters in unimodular $f(R,T)$ gravity, including the scalar spectra index ($n_s$), we use the power spectrum which is defined as follows
\begin{equation}
{\cal{P}}_s(k)\approx\frac{H^2}{8\pi^2 M_{p}^2{c_s^3}\varepsilon_1 }\bigg|_{c_s k=aH}\,.
\end{equation}
This relation is evaluated at the horizon crossing time for which $c_sk = aH$ with $k$ being the wave number. The sound speed for this model is $c_s^2=1$. By using the equations (25) and (31) we obtain
\begin{equation}
{\cal{P}}_s(k)\approx\frac{(\kappa^2+2\beta-4\alpha_2)(\kappa^2+\beta)}{12\pi^2 M_{p}^2(1-4\alpha_1)}\frac{V^3(\phi)}{V^{\prime 2}(\phi)}\bigg|_{c_s k=aH}\,\,.
\end{equation}
Finally we arrive at the following relation for the scalar spectral index as a function of the scalar field potential
\begin{equation}
1-n_s\approx 2\varepsilon_1\Bigg[1+4\bigg(1-\frac{V^{\prime\prime}(\phi)V(\phi)}{V^{\prime 2}(\phi)}\bigg)\Bigg]\,.
\end{equation}

This result will be used in forthcoming arguments.

\section{Reheating in unimodular $f(R,T)$ gravity}

 In the cold inflation paradigm (in contrast to the warm inflation), at the end of inflation the universe is cold and dark. In fact, in cold inflation the universe is more or less empty of particles at the end of inflation and just a nearly homogenous inflaton field is dominated. To have other species of matter such as the standard model particles and all relevant degrees of freedom, the energy of inflaton field transferred to feed processes such as the Nucleosynthesis. So, in the reheating period, the inflaton field converts its energy to relativistic particles during its coherent rapid oscillation about the minimum of the potential. The interaction of these particles leads to thermal equilibrium of universe which is characterized by the reheating temperature $T_{reh}$, and then universe entered the radiation dominated era. To determine the reheating temperature, we follow the method in Refs.~\cite{Goo18, Mie11}. To this end, we consider the universe expansion between the time of the horizon crossing in inflation era and the time of re-entering. Therefore, the e-folds number is given as follows
\begin{eqnarray}
N=\ln{\frac{a_0}{a_*}}=\ln{\bigg((\frac{a_{end}}{a_*})(\frac{a_{reh}}{a_{end}})(\frac{a_{rec}}{a_{reh}})(\frac{a_{0}}{a_{rec}})\bigg)} \\ \nonumber
=N_1+N_2+N_3+N_4\,,\quad\quad\quad\quad\,\,\,\,
\end{eqnarray}
which is divided into four sub-eras: 1- The inflation era from the time of horizon crossing until the end of inflation ($t_{end}$). 2- Rapid oscillation from the end of inflation until reheating or begining of the radiation dominated era ($t_{reh}$). 3- From reheating until the recombination era ($t_{rec}$) and finally from recombination era until the present time ($t_0$) where e-folds numbers denote the value of this parameter at their corresponding sub-era. Now, we calculate the e-folds numbers for each sub-era. By using the equation (33) we obtain the number of e-folds $N_1$ for the potential $V(\phi)=q\,\phi^n$ as
\begin{equation}
N_1=\frac{(\beta+\kappa^2)}{2n} (\phi_{*}^{2}-\phi_{end}^{2})\,.
\end{equation}
The spectral index (43) for power law potential is given by
\begin{equation}
1-n_s\approx \frac{n^2+4n}{(\beta+\kappa^2)\phi_*^2}\,.
\end{equation}
Using this equation we rewrite the e-folds number (45) as follows
\begin{equation}
N_1= \frac{n+4}{2(1-n_s)}-\frac{n}{4}\,,
\end{equation}
and we obtain power spectrum at the horizon crossing as
\begin{equation}
{\cal{P}}_s(k)\approx\frac{(\kappa^2+2\beta-4\alpha_2)(\kappa^2+\beta)}{12\pi^2 M_{p}^2(1-4\alpha_1)}\,\,\frac{q\,\phi_{*}^{n+2}}{n^2}\,.
\end{equation}
The Hubble parameter at the horizon crossing becomes
\begin{eqnarray}
H^2\approx 8\pi^2M_p^2c_s^3 \varepsilon_1 {\cal{P}}_s(k)\bigg|_{c_s k=aH}
\approx 4\pi^2 M_p^2 \,\frac{n(1-n_s)}{n+4}{\cal{P}}_s(k)\bigg|_{c_s k=aH} \,.
\end{eqnarray}
At the end of inflation era, we have $\varepsilon_1\approx1$. So, we find
\begin{equation}
\phi_{end}\approx\frac{n}{\sqrt{2(\beta+\kappa^2)}}\,.
\end{equation}
The energy density in this region is
\begin{equation}
\rho_{end}\approx V(\phi_{end})\approx q\,\phi_{end}^n\approx q\bigg[\frac{n^2}{2(\beta+\kappa^2)}\bigg]^{\frac{n}{2}}\,.
\end{equation}
After the slow roll, inflaton field oscillates rapidly around the minimum of the potential. Following the Ref.~\cite{Han04}, the behavior of quasiperiodic oscillation of the inflaton field in this case can be defined as
\begin{eqnarray}
\phi=\Phi(t) \sin\Big(\int{W(t)dt}\Big)\,,
\end{eqnarray}
where $W(t)$ is a function of the cosmic time and $\Phi(t)$ is a time dependent amplitude given as $V(\Phi(t))=q\Phi^{n}(t)=\rho_{\phi}$. Note that $|\frac{\dot{\Phi}}{\Phi}|\ll W(t)$ in the high frequency oscillation after slow roll. Note also that this implies that the energy density and the Hubble parameter during this high frequency oscillation decrease very
slowly in a period of oscillation of the scalar field, see for instance~\cite{Sadjadi2013}.

The equation of state parameter of the scalar field, $\omega_{\phi}$, is given by
\begin{eqnarray}
\gamma=1+\omega_\phi=\frac{<{\rho_\phi+p_\phi}>}{<\rho_\phi>}=\frac{<\dot{\phi}^2>}{<\rho_{\phi}>} \nonumber\\
=\frac{2<\rho_\phi-V(\phi)>}{V(\Phi)}=\frac{2n}{n+2}\,,
\end{eqnarray}
where $<f>$ means time average of the quantity $f$. The energy density during the reheating era is given by $\rho \sim a^{-3(1+\omega_\phi)}$. Hence, the number of e-folds in the reheating era is written as
\begin{eqnarray}
N_2=\ln\Big(\frac{a_{reh}}{a_{end}}\Big)=-\frac{1}{3(1+\omega_\phi)}\ln\Big(\frac{\rho_{reh}}{\rho_{end}}\Big)\,.
\end{eqnarray}
The relation between energy density and temperature of reheating is expressed as ~\cite{Coo15, Cai15}
 \begin{eqnarray}
\rho_{reh}=\frac{\pi^2 g_{reh}}{30}T_{reh}^4\,,
\end{eqnarray}
where $g_{reh}$ is the effective number of the relativistic species in reheating era. Using equations (51) and (55), we derive the e-folds number during rapid oscillation as follows
\begin{eqnarray}
N_2=-\frac{1}{3(1+\omega_\phi)}\ln\Bigg(\frac{\frac{g_{reh}}{30}\,\pi^2 \,T_{reh}^4}{q\,\big(\frac{n^2}{2(\beta+\kappa^2)}\big)^{\frac{n}{2}}}\Bigg)\,.
\end{eqnarray}
By using the equation (46), we find the scalar field at the horizon crossing as
\begin{eqnarray}
\phi_*\approx \bigg(\frac{n^2+4n}{(\beta+\kappa^2)}\frac{1}{1-n_s}\bigg)^{1/2}\,.
\end{eqnarray}
Also, one can obtain $q$ from equation (48) as
\begin{eqnarray}
q\approx\frac{12\pi^2M_p^2\,n^2(1-4\alpha_1){\cal{P}}_s(k_0)}{(\kappa^2+2\beta-4\alpha_2)(\kappa^2+\beta)\,(1-n_s)^{-\frac{n+2}{2}}}
\bigg(\frac{n^2+4n}{\beta+\kappa^2}\bigg)^{-\frac{n+2}{2}}\,.
\end{eqnarray}
Eventually, the e-folds number (56) becomes
\begin{eqnarray}
N_2=-\frac{1}{3(1+\omega_\phi)}\ln\Bigg(\frac{\frac{g_{reh}}{30}\,\pi^2\, T_{reh}^4{(1-n_s)^{\frac{n+2}{2}}}}{{24\pi^2M_p^2{\cal{P}}_s(k_0)}\frac{(1-4\alpha_1)}{(\kappa^2+2\beta-4\alpha_2)}\big(\frac{n}{2(n+4)}\big)^{\frac{n}{2}+1}}\Bigg)\,.
\end{eqnarray}
In the next step, we calculate the number of e-folds in recombination era. After the rapid oscillation, the universe is composed of ultra-relativistic particles in thermal equilibrium. Also, it undergoes an adiabatic expansion. In this case, we have $s=Sa^{-3}$ where the entropy density is defined as ~\cite{Mie11, Goo18}
\begin{eqnarray}
s=\frac{2\pi^2}{45}g_{reh}T^3\,.
\end{eqnarray}
Therefore, we can write
\begin{eqnarray}
\frac{a_{rec}}{a_{reh}}=\frac{T_{reh}}{T_{rec}}\Big(\frac{g_{reh}}{g_{rec}}\Big)^{\frac{1}{3}}\,.
\end{eqnarray}
The only degree of freedom in this era is photon with $g_{rec}=2$. So, we have
\begin{eqnarray}
N_3=\ln\bigg(\frac{T_{reh}}{T_{rec}}\Big(\frac{g_{reh}}{2}\Big)^{\frac{1}{3}}\bigg)\,.
\end{eqnarray}
The temperature is related to redshift as $T(z)=T(z=0)(1+z)$.
So, we can write the reheating temperature in terms of the present CMB temperature
\begin{eqnarray}
T_{rec}=(1+z_{rec})T_{CMB}\,.
\end{eqnarray}
Then, we obtain
\begin{eqnarray}
N_3+N_4=\ln\bigg(\frac{T_{reh}}{T_{CMB}}\Big(\frac{g_{reh}}{2}\Big)^{\frac{1}{3}}\bigg)\,.
\end{eqnarray}
Now, for computing the reheating temperature we use the relation (44). By using the equation (49) and putting $a_0=1$, the right hand side of equation (44) can be written as
\begin{eqnarray}
\ln{\frac{a_0}{a_*}}=\ln{\frac{H_*}{k_0}}\approx \ln{\Bigg(\frac{2\pi M_p}{k_0}\sqrt{\frac{n(1-n_s)}{n+4}{\cal{P}}_s(k_0)}\Bigg)}\,.
\end{eqnarray}
Finally, by substituting equations (47), (59), (64) and (65) into equation (44) we obtain the reheating temperature as
\begin{eqnarray}
T_{reh}\approx A \,g_{reh}^{m_8}\, M_p^{2m_2}\,\Big(\frac{T_{CMB}}{k_0}\Big)^{m_1}\,,
\end{eqnarray}
where
\begin{eqnarray}
A=\exp\Bigg(\frac{3n^{2}}{4(n-4)}\Big(1-\frac{2(n+4)}{n(1-n_{s})}\Big)\Bigg)
\times \Bigg(\frac{2^{m_3}\pi^{m_1}\,n^{m_4}(n+4)^{m_5}\Big(\frac{\kappa^2+2\beta-4\alpha_2}{1-4\alpha_1}\Big)^{m_7}}{(45)^{m_7}\big({\cal{P}}_s(k_0)\big)^{-m_2}(1-n_s)^{-m_6}}\Bigg)\,.
\end{eqnarray}
and $m_i$'s are defined as
\begin{eqnarray}
m_1=\frac{3n}{n-4}\,, \hspace{1.6cm}
m_2=\frac{n-1}{n-4}\,,\hspace{1.6cm}
m_3=\frac{(n-2)^2}{4(n-4)}\,,\hspace{1.1cm}\nonumber\\
m_4=-\frac{n^2-2n+4}{4(n-4)}\,,\hspace{0.4cm}
m_5=\frac{n^2-2n+4}{4(n-4)}\,, \hspace{0.5cm}
m_6=\frac{n^2+10n+4}{4(n-4)}\,,\hspace{0.3cm} \nonumber\\
m_7=\frac{n+2}{2(n-4)}\,, \hspace{2.1cm}
m_8=\frac{2-n}{2(n-4)}\,. \hspace{4cm}
\end{eqnarray}
Equation (66) shows that the reheating temperature can be written in terms of the CMB temperature, the spectral index, the power spectrum and the parameters of the model. If we consider $\beta=\alpha_1=\alpha_2=0$ the reheating temperature for quadratic potential is as follows
\begin{eqnarray}
T_{reh}\approx 2.0524 M_p\sqrt{\frac{1-n_s}{{\cal{P}}_s}}\Big(\frac{k_0}{T_{CMB}}\Big)^3 e^{\frac{6}{1-n_s}}\,.
\end{eqnarray}
This relation is the same as the reheating temperature of standard general relativity ~\cite{Mie11}. To have some numerical estimations, for the quadratic potential and taking the constant parameters as $T_{CMB}=2.725 K=2.348\times10^{-4} eV$, $g_{reh}=106.75$, $k_0=0.002M pc^{-1}$ (with $Mpc^{-1}=6.39\times 10^{-30}eV$) and by using, for instance,  WMAP7 data ~\cite{Kom11} as
\begin{eqnarray}
{\cal{P}}_s(k_0)=2.441^{+0.088}_{-0.092}\times 10^{-9},\\\nonumber
n_s=0.963\pm 0.012,\hspace{1.5cm}
\end{eqnarray}
we find the reheating temperature as $T_{reh}=2.125\times 10^{12}$ GeV which is larger than what was obtained in standard general relativity (about 2 order of magnitude).
For a comparison of the obtained reheating temperature with other estimates in some other inflation models, we note that an allowed range of variation for
the reheating temperature in large field inflation with $g_{reh}\simeq 200$ has been obtained in \cite{Mar10} as $6$ TeV$\lesssim T_{reh}\lesssim 10^{4}$ TeV.
For a perfect fluid model with an exponential form, the reheating temperature is restricted to $T_{reh}\simeq 10^{15}$GeV~\cite{Mirtalebian21}. For a special constant-roll inflation this temperature could be of the order of $\sim 10^{15}$ GeV \cite{Sadjadi2020}. For a modified teleparallel model of inflation, $T_{reh}=6\times 10^{14}$GeV~\cite{Goo18}. Recently Passaglia et al. \cite{Passaglia2021} via a Higgs condensate mechanism have achieved the highest temperature during reheating as $T_{reh}\sim 10^{7}$GeV $ \Big(\frac{H_{end}}{10^{10}GeV}\Big)$, where $H_{end}$ is the Hubble rate at the end of inflation. With $H_{end}\lesssim 10^{11} GeV$, one obtains $T_{reh}\sim 10^{8} GeV$ as the maximum temperature for reheating in their model. There are further numerical estimations for reheating temperature in variety of inflation models that can be seen in literature.

\section{Gravitational particle creation in unimodular $f(R,T)$ gravity}

In this section, we study the gravitational particle creation in the unimodular $f(R,T)$ gravity as a modification of general relativity. The problem of particle production in expanding universe has been discussed by several authors; see for instance~\cite{Par69, Par71, Gri05, Gri93, Bat08, Pav08}. To study the particle creation in a modified gravity, we consider quantization in curved background ~\cite{Bir82, Muk07}. It is known that any extended theory such as $f(R,T)$ gravity can be recast as general relativity with some non-minimally coupled field(s)~\cite{Cap15}. In order to calculate the rate of particle production, we study the Bogolubov transformations in the context of homogeneous and isotropic cosmologies. To this end, we consider a minimally coupled massless scalar field, which describes the created particles, in the matter action as
\begin{eqnarray}
S=\frac{1}{2\kappa^2}{\int}{d^4x\Big[\sqrt{-g}f(R,T)-2\lambda(\sqrt{-g}-1)\Big]}
+\frac{1}{2}{\int}d^4x\sqrt{-g}{g^{\mu\nu}\partial_{\mu} \phi\,\partial_\nu \phi}\,.\nonumber\\
\end{eqnarray}
The equation of motion is given by varying of the action with respect to the field
\begin{eqnarray}
-g^{\mu\nu}\phi_{,\mu\nu}-\frac{1}{\sqrt{-g}}(g^{\mu\nu} \sqrt{-g})=0\,,
\end{eqnarray}
where a comma marks spacetime derivatives, e.g. $\partial_{\mu}\phi \equiv \phi_{,\mu}$. The trace of the energy-momentum tensor is
\begin{eqnarray}
T_\mu^{\mu}=-g^{\mu\nu}\partial_{\mu} \phi\,\partial_\nu {\phi}=-(\partial \phi)^2\,.
\end{eqnarray}
A spatially flat FRW spacetime is also a conformally flat spacetime. In order to explicitly transform the FRW metric into a conformally flat form, we introduce the conformal time $\eta$ which is defined as $\eta(t)\equiv \int_{t_0}^{t}{\frac{dt}{a(t)}}$. Thus, the line element gets the following form
\begin{eqnarray}
ds^2=a^2(\eta)(-d\eta^2+dX^2)\,,
\end{eqnarray}
where $g_{\mu\nu}=a^2 \eta_{\mu\nu}$ and $\sqrt{-g}=a^4$. In the coordinates $(\eta,x)$ and by definition $\chi(\eta,x) \equiv a(\eta)\phi(\eta,x)$, the action of the scalar field becomes
\begin{eqnarray}
S=\frac{1}{2} \int{d^3\textbf{x} d\eta \Big[\chi'^2-(\nabla \chi)^2+\frac{a''}{a}\chi^2\Big] }\,.
\end{eqnarray}
The equation of motion for $\chi(\eta, \textbf{x})$ can be written as
\begin{eqnarray}
3f''_{,R}(R,T)+6\frac{a'}{a}f'_{,R}(R,T)-a^2f_{,R}(R,T)R+2a^2f(R,T)-4a^2\lambda\\\nonumber=-\frac{\kappa^2+3f_{,T}(R,T)}{a^2}\Big[\,\chi'^2-(\vec{\nabla} \chi)^2+\frac{a'^2}{a^2}\chi^2-\frac{a'}{a}(\chi\chi'+\chi'\chi)\Big]\,,
\end{eqnarray}
with
\begin{eqnarray}
R=6\frac{a''}{a^3}\,,
\end{eqnarray}
and
\begin{eqnarray}
\chi''-\Delta \chi+\frac{1}{6}a^2 R \chi=0\,,
\end{eqnarray}
where a prime denotes derivative with respect to the conformal time $\eta$ and $\Delta$ is the Laplacian. We note that the dynamics of the scalar field $\phi$ in a flat FRW spacetime and the auxiliary field $\chi$ in the Minkowski spacetime are mathematically equivalent. The action (75) is explicitly time dependent, so the energy of the field $\chi$ is generally not conserved which leads to the possibility of particle creation in quantum theory. The energy of new particles is provided by the gravitational field. By expanding the field $\chi$ in terms of Fourier modes as
\begin{eqnarray}
\chi(\eta,\textbf{x})=\int{\frac{d^3\textbf{k}}{(2\pi)^\frac{3}{2}}}\chi_{\textbf{k}}(\eta)e^{i\textbf{k}.\textbf{x}}\,,
\end{eqnarray}
where $\textbf{k}$ is the wave vector, $\textbf{k}\equiv (k_1, k_2, k_3)$, we obtain the equation of motion as follows
\begin{eqnarray}
\chi''_k+\omega_k^2(\eta)\chi_k=0\,,
\end{eqnarray}
with
\begin{eqnarray}
\omega_k=\sqrt{k^2-\frac{1}{2}a^2 R}\,.
\end{eqnarray}
Equation (80) is analogous to the harmonic oscillator with $\omega_{\textbf{k}}$ depending on conformal time $\eta$. The solution of this equation gives positive and negative frequencies modes. If the quantum field was in a vacuum state at $t=-\infty$($\eta=-\infty$), then $\chi_k=\exp(-ik\eta)$. The general solution can be represented in the form of an integral equation as a complete set of mode-solutions for the field $\chi(\eta,\textbf{x})$ as follows
\begin{eqnarray}
\chi(\eta,\textbf{x})=\frac{1}{(2\pi)^{\frac{3}{2}}}\frac{1}{\sqrt{2}}\int{d^3\textbf{k}}\big[a_k^- \chi_k^*(\eta)e^{\textbf{ik.x}}+a_k^+ \chi_k(\eta)e^{-\textbf{ik.x}}\big]\,,
\end{eqnarray}
where $a_k^-$ and $a_k^+$ are respectively the annihilation and creation operators. We can express $\chi(\eta)$ as
\begin{eqnarray}
\chi_k(\eta)=\frac{1}{k}\int_0^\eta \sin{k(\eta-\eta')}V(\eta')\chi(\eta')d\eta'+e^{ik\eta}\,,
\end{eqnarray}
where
\begin{eqnarray}
V(\eta)=\frac{1}{6}a^2 R\,.
\end{eqnarray}
Since $t\rightarrow +\infty$ ($\eta\rightarrow+\infty$), we have $\chi_k(\eta)=\alpha_k e^{-ik\eta}+\beta_k e^{ik\eta}$ where $|\alpha_k|^2-|\beta_k|^2=1$. $\alpha_k$ and $\beta_k$ are the Bogolyubov coefficients
\begin{eqnarray}
\alpha_k=1+\frac{i}{2k}\int_{-\infty}^{\infty}e^{ik\eta}\,V_k(\eta)\,\chi_k(\eta)\,d\eta\,,
\end{eqnarray}
and
\begin{eqnarray}
\beta_k=-\frac{i}{2k}\int_{-\infty}^{\infty}e^{-ik\eta}\,V_k(\eta)\,\chi_k(\eta)\,d\eta\,.
\end{eqnarray}
The expectation value of the particle number operator $N_k=a_k^+ a_k^-$ can be obtained as ~\cite{Muk07}
\begin{eqnarray}
<0|N|0>=|\beta_k|^2\,,
\end{eqnarray}
The total density of the real produced particles is obtained as
\begin{eqnarray}
n=\frac{1}{(2\pi)^3 a^3}\int{d^3 \textbf{k}} |\beta_k|^2\,.
\end{eqnarray}
By solving equation (83) and substituting $\chi(\eta)$ in equation (86) we find
\begin{eqnarray}
n(t)=\frac{1}{576\pi a^3}\int_{-\infty}^{\infty} dt' a^3 R^2\,.
\end{eqnarray}
Now, we assume the power-law scale factor as $a(t)=bt^m$ where $b$ and $m$ are positive, real numbers. We consider the specific type of $f(R,T)=f_1(R)+f_2(T)$ where $f_1(R)$ and $f_2(T)$ are arbitrary functions of their arguments. By substituting this scale factor in the field equations we obtain
\begin{eqnarray}
f_1(R)=B_+ R^{\mu_{+}+1}+B_{-}R^{\mu_{-}+1}+DR^{\frac{3}{2}m}
\end{eqnarray}
where $\mu_{\pm}=-\frac{m+1\pm\sqrt{m^2+10m+1}}{4}$ and
\begin{eqnarray}
B_{\pm}=C_{\pm}[b^3(3m+1)]^{\frac{2\mu_{\pm}}{3m+1}}\frac{1}{(\mu_{\pm}+1)(12m^2-6m)^{\mu_\pm}}\,,
\end{eqnarray}
\begin{eqnarray}
D=\frac{2}{3m}A[b^3(3m+1)]^{\frac{3m-2}{3m+1}}\frac{1}{(12m^2-6m)^{\frac{3}{2}m-1}}\,,\hspace{1.1cm}
\end{eqnarray}
where $A$ and $C_{\pm}$ are integration constants and

\begin{eqnarray}
f_2(T)=-A\,b^{-\frac{9}{3m+1}}(3m+1)^{\frac{3m-2}{3m+1}}\Big(\frac{T}{3\omega_m-1}\Big)^{\frac{1}{1+\omega_m}}-\kappa^2 T\,.
\end{eqnarray}
Finally the unimodular Lagrange multiplier, $\lambda$, is obtained as
\begin{eqnarray}
\lambda(\tau)=N_+\,\tau^{\frac{-2(\mu_{+}+1)}{3m+1}}+N_-\,\tau^{\frac{-2(\mu_{-}+1)}{3m+1}}+N_1 \tau^{\frac{-3m}{3m+1}}+N_2 \tau^{\frac{-3(1+\omega_m)m}{3m+1}}\,,
\end{eqnarray}
where
\begin{eqnarray}
N_{\pm}=\Big\{m(1-3m)+\frac{3m(2m-1)}{\mu_{\pm}+1}-10m\mu_{\pm}+2(2\mu_{\pm}+3m+1)\Big\}\Big[b^3(3m+1)\Big]^{-\frac{2m}{3m+1}}\,,
\end{eqnarray}

\begin{eqnarray}
N_1=\frac{(1-3\omega_m)}{2}A(12m^2-13m+2)\Big[b^3(3m+1)\Big]^{-\frac{2m}{3m+1}}\,,\hspace{4.8cm}
\end{eqnarray}
and
\begin{eqnarray}
N_2=\frac{(1-3\omega_m)}{2}\kappa^2\Big[\frac{b^{\frac{1}{m}}}{3m+1}\Big]^{\frac{3(1+\omega_m)m}{3m+1}}\,.\hspace{7.8cm}
\end{eqnarray}
Equation (94) shows that in this scenario the Lagrange multiplier $\lambda$, which plays the role of a cosmological constant in unimodular viewpoint, is varying with time. This is an interesting result since an evolving cosmological constant provides some new facilities for the rest of cosmology, especially for the late time cosmic dynamics.  The scale factor $a(t)$ in terms of the conformal time $\eta$ becomes
\begin{equation}
a(\eta)=\frac{B}{(-\eta)^{\frac{m}{m-1}}}\,, \quad\quad B=[b(m-1)^m]^{\frac{1}{1-m}}\,, \quad -\infty<\eta<0\,.
\end{equation}
These solutions can be regarded as approximations to more realistic models such as the standard fluid (like radiation dominated universe and dust) and provide a framework for establishing the behavior of more general cosmological solutions. To determine the parameter $b$, we assume the scale factor is normalized to unity for the present time, $a(t_0)\equiv a_0=1$. The Hubble parameter, $H=\frac{m}{t}$ roughly approximates the age of the universe as $t_0=\frac{m}{H_0}$, that is of the order of $10^9$ years. So we get $b=(\frac{H_0}{m})^m$. Now, by inserting the scale factor into equations (80), we find the mode function as follows
\begin{eqnarray}
\chi_k''+\bigg[k^2-\bigg(\frac{m(2m-1)}{(m-1)^2}\bigg)\frac{1}{\eta^2}\bigg]\chi_k=0\,,
\end{eqnarray}
where $\omega_k^2(\eta)\equiv k^2-\bigg(\frac{m(2m-1)}{(m-1)^2}\bigg)\frac{1}{\eta^2}$. This is a second order differential equation with two independent solutions. For all times, these modes must be normalized according to the Wronskian
\begin{eqnarray}
W_k(\eta)\equiv\chi_{1k}(\eta)\chi'_{2k}(\eta)-\chi'_{1k}(\eta)\chi_{2k}(\eta)=-2i\,.
\end{eqnarray}
Also they must satisfy the initial conditions at the time $\eta_i$
\begin{eqnarray}
\chi_k(\eta_i)=\frac{1}{\sqrt{\omega_k(\eta_i)}}\,,\qquad\quad \chi'(\eta_i)=i\sqrt{\omega_k(\eta_i)}.
\end{eqnarray}
The general solution of equation (99) which satisfies the initial conditions (101) and the normalization condition (100) is given as follows
\begin{eqnarray}
\chi_{k}(\eta)=A(k,\eta_i) J_{\nu}(k\eta)-B(k,\eta_i) Y_{\nu}(k\eta)\,,\qquad \nu\equiv \sqrt{\frac{1}{4}+\frac{m(2m-1)}{(m-1)^2}}\,,
\end{eqnarray}
where $J_\nu$ and $Y_\nu$ are the Bessel functions of the first and second kind respectively and $\nu$ is the order of these functions. $A$ and $B$ are complex temporally constants, depending on $k$ and the initial conformal time $\eta_i$
\begin{eqnarray}
A(k,\eta_i)=\frac{2k\eta_i Y_{\nu+1}(k\eta_i)-2\nu Y_{\nu}(k\eta_i)-Y_{\nu}(k\eta_i)+2I\eta_i\omega_k(\eta_i)Y_{\nu}(k\eta_i)}{2k\eta_i^{3/2}\sqrt{\omega_k(\eta_i)\Big[J_\nu(k\eta_i)Y_{\nu+1}(k\eta_i)-J_{\nu+1}(k\eta_i)Y_{\nu}(k\eta_i)\Big]}}\,,
\end{eqnarray}
\begin{eqnarray}
B(k,\eta_i)=\frac{2k\eta_i J_{\nu+1}(k\eta_i)-2\nu J_{\nu}(k\eta_i)-J_{\nu}(k\eta_i)+2I\eta_i\omega_k(\eta_i)J_{\nu}(k\eta_i)}{2k\eta_i^{3/2}\sqrt{\omega_k(\eta_i)\Big[J_\nu(k\eta_i)Y_{\nu+1}(k\eta_i)-J_{\nu+1}(k\eta_i)Y_{\nu}(k\eta_i)\Big]}}\,.
\end{eqnarray}
Following Ref.~\cite{Muk07}, a straightforward calculation leads to the final expression for the number density of particles created in the $k$ mode at time $\eta>\eta_i$ as
\begin{eqnarray}
N_k(\eta)=\frac{1}{4\rvert\omega_k(\eta)\lvert}\rvert\chi'_{k}\lvert^2+\frac{\rvert\omega_k(\eta)\lvert}{4}\rvert\chi_{k}\lvert^2-\frac{1}{2}\,.
\end{eqnarray}
This equation shows that when $k^2=\frac{m(2m-1)}{(m-1)\eta^2}$, the frequency vanishes and the number density of created particles diverges. Fortunately, such divergence is removed when the integration of $k$ starts from $k_{min}$, as must be in our case. Thus, the integral of the total particle number is finite. The negative value of frequency expresses that the state of minimum energy and the quantum vacuum are not well-defined and the particle production ceases for these values. The final form of $N_k(\eta)$ is a quite complicated function depending on the initial conformal time $\eta_i$, $k$ mode and on the parameter $m$. We can use $\eta\equiv\int{\frac{dt}{a(t)}}$ to recover the spectrum in terms of the physical time $t$ and also all quantities can be expressed in terms of $t$. Now, we study the time evolution of the total particle number $N_k$. To this end, we consider an appropriate time scale. The present time corresponds to $t_0\sim H_0^{-1}=4.4\times 10^{17} s$, so an appropriate time scale is in terms of $H_0^{-1}$ .
To establish the value of the initial time $t_i$, we assume a universe dominated by matter and we know (from Ref.~\cite{Kob90}) that the matter and radiation decouple at about $t_i=10^{13} s\approx4.5\times 10^{-5}H_0^{-1}$; we will use such a value for the initial time. Thus we are interested
in the interval $4.5\times 10^{-5}H_0^{-1}<t<H_0^{-1}$, which corresponds to the matter dominated era. In figure 2 we have shown the behavior of massless created particles' density for three different wave numbers $k$. We can see that for each $k$ there is an abrupt growth of the particle number at the initial time and after a brief oscillation it turns constant and remains so throughout evolution. This figure shows also that for smaller values of $k$ the amount of created particles are increasing. This feature indicates that higher modes contribute much less to the total number of particles. We note also that suppression of gravitational particle creation at late time may be as a result of free streaming effects governed by the dark matter mass. Also, as has been shown in~\cite{Sarkar15}, the suppression may be controlled also by the redshift of late forming dark matter. 
\begin{figure*}
\begin{center}	{\includegraphics[width=.46 \textwidth,origin=c,angle=0]{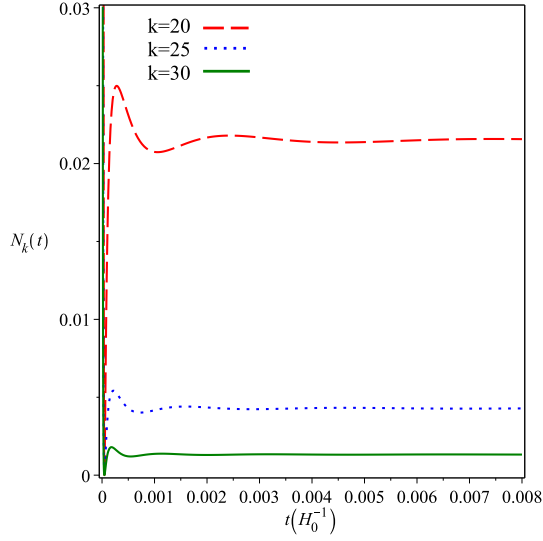}}
\end{center}
\caption{\small{Evolution of the density of massless created particles $N_k$ versus the physical time $t$ (in units of $H_{0}^{-1}$) for three different values of the mode $k$ with the parameter $m=3/5$ which corresponds to $\nu=1$.}}
\end{figure*}

Now, we consider $m = \frac{2}{3}$ which corresponds to the
dust domination case. Then, the normalized mode function takes the following form
\begin{eqnarray}
\chi_k(\eta)=A(k,\eta_i) \big(1+\frac{i}{k\eta}\big)e^{ik\eta}-B(k,\eta_i) \big(1-\frac{i}{k\eta}\big)e^{-ik\eta}\,,\qquad
\end{eqnarray}
where
\begin{eqnarray}
A(k,\eta_i)=-\frac{(-k^2\eta_i^2 +ik\eta_i-\omega(\eta_i)k\eta_i^2 +i\omega(\eta_i)\eta_i+1)e^{-ik\eta_i}}{2k^2\eta_i^2\sqrt{\omega(\eta_i)}}\,,
\end{eqnarray}
\begin{eqnarray}
B(k,\eta_i)=\frac{(k^2\eta_i^3\omega(\eta_i)-k^3\eta_i^3+\omega(\eta_i)\eta_i-i )e^{ik\eta_i}}{2k^2\eta_i^2\sqrt{\omega(\eta_i)}(-k\eta_i+i)}\,. \hspace{2cm}
\end{eqnarray}
\begin{figure*}
\begin{center}	{\includegraphics[width=.46 \textwidth,origin=c,angle=0]{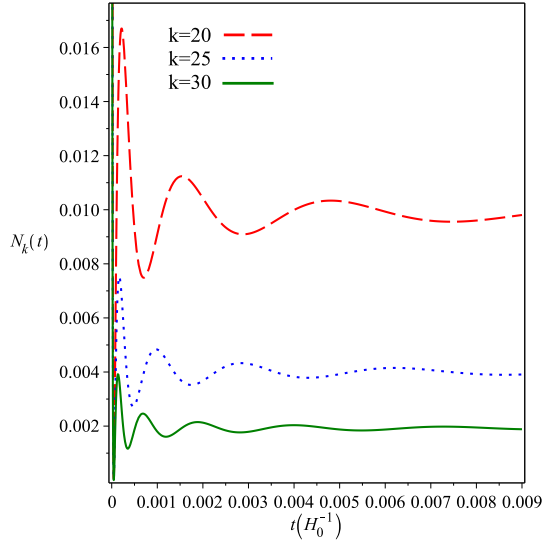}}
\end{center}
\caption{\small{Evolution of the density of massless created particles $N_k$ versus the physical time $t$ (in units of $H_{0}^{-1}$) for three different values of the mode $k$ with the parameter $m=2/3$.}}
\end{figure*}
In figure 3 we have depicted this function for some values of parameter $k$. We see that for larger values of $m$, the initial oscillations increases. We also note that the density of massless created particles decreases for higher values of the parameter $m$.

Now, we study thermodynamics of irreversible processes in the presence of gravitationally created particles. Non-conservation of the matter stress-energy tensor $\nabla^{\mu}T_{\mu\nu}\neq 0$, implies a transfer of energy from the geometry to the matter sector\cite{Har14,Har15,Har20}. In fact, non-conservation of the matter stress-energy tensor in related theories can describe particle creation on a cosmological scale. To this end, we use the formalism of irreversible thermodynamics of open systems in the presence of matter creation. Let us consider the universe treated as an open thermodynamic system allowing irreversible matter creation from the energy of the gravitational field. Thus, the creation of matter acts as a source of internal energy. In such a situation, the number of particles $N$ in a given volume $V=a^3$ is not a constant, but is time dependent. For this cosmological system, the thermodynamical conservation equation written in its most general form, is given by~\cite{Pri86,Lob22}
\begin{eqnarray}
d(\rho V)=dQ-pdV+\frac{h}{n}d(nV)\,,
\end{eqnarray}
where $dQ$ is the heat received by the system during time $dt$, $n = N/V$ is the particle number density and $h=\rho +p$ is the enthalpy per unit volume. We restrict our analysis to adiabatic transformations defined by the condition $dQ = 0$. Therefore, we ignore proper heat transfer processes in the cosmological system. So, we can reformulate equation (109) in an equivalent form as
\begin{eqnarray}
\dot{\rho}+3(\rho+p)H=\frac{\rho+p}{n}(\dot{n}+3Hn)\,.
\end{eqnarray}
We see that in such a system the thermal energy is received due to the change in the particle number. We introduce the particle creation rate as
\begin{eqnarray}
\Gamma=\frac{\dot{n}}{n}+3H\,,
\end{eqnarray}
Then, by comparing equations (17) and (110), we find that the particle creation rate is as follows
\begin{eqnarray}
\Gamma=-\frac{1}{(\kappa^2+f_{,T})(\rho+p)}\Big[-\frac{1}{2}\dot{T}f_{,T}+(\rho_m+p_m)
\dot{f}_{,T}+\dot{p}_{m}f_{,T}+\dot{\lambda}\Big]\,.
\end{eqnarray}
Therefore, the energy conservation equation can be reformulated in the alternative form as
\begin{eqnarray}
\dot{\rho}+3(\rho+p)H=(\rho+p)\Gamma\,.
\end{eqnarray}
For adiabatic transformations describing irreversible
particle creation in an open thermodynamic system, the
first law of thermodynamics can be rewritten as an effective
energy conservation equation as
\begin{eqnarray}
\dot{\rho}+3(\rho+p+p_c)H=0\,,
\end{eqnarray}
where we have defined a new thermodynamic quantity, $p_c$, which denotes the creation pressure as
\begin{eqnarray}
p_{c}=-\frac{\rho+p}{3}\frac{\Gamma}{H}\,.
\end{eqnarray}
By solving the differential equation (6) we obtain the particle number density as
\begin{eqnarray}
n=t^{-3m}e^{\int{\Gamma(t)dt}}\,.
\end{eqnarray}

To have some numerical estimation of the density of the newly created particles, following \cite{Hashiba2019}, we consider a smooth transition from cosmological inflation to kination at the time scale $\Delta t$. In this situation, the energy density of produced particles is given by $\rho= {\cal {C}} e^{-4m\Delta t} m^{2} H_{inf}^{2}$ where ${\cal {C}}\simeq 2\times 10^{4}$ and $m$ and $H_{inf}$ denote mass of the created particle and Hubble parameter during inflation. If we consider two massive particles, their mass
terms serve as a source of gravitational particle production. One of them decays
into radiation and realizes reheating, and the other is a stable purely gravitational dark matter. We call them $A$ and $X$, respectively. In Ref.~\cite{Hashiba2019}, it has been shown that $m_{A}\simeq (\Delta t)^{-1}\simeq H_{inf}\simeq 10^{13}$ GeV. Therefore, to explain sufficient reheating and also the present dark matter abundance at
the same time, $m_{A}$ should be of the order of $10^{13}$GeV and $m_{X}\sim10^{3}$GeV.

\section{Reheating in Einstein frame}

In this section we analyze the reheating process in the Einstein frame. We can express unimodular $f(R,T)$ gravity action in terms of the scalar degree of freedom by redefinition of the variables of the model. The conformal transformation of the action from Jordan frame to Einstein frame makes the scalar degree of freedom more explicit and enables us to use the analogy of the scalar field inflation. Hence, we use the analogy of two-fields inflation. Since we regard the Jordan frame as the physical frame, we need to recast resulting quantities obtained in the Einstein frame back to the Jordan frame after the calculation. We can recast this theory to the Einstein gravity with the scalar fields by introducing the following conformal transformation
\begin{equation}
\tilde{g}_{\mu\nu}=\Omega^2 g_{\mu\nu}\,,
\end{equation}
where $\Omega^2$ is the conformal factor. Note that the quantities in the Einstein frame are represented by a
tilde over them. We define the canonical scalar fields as
\begin{equation}
\Omega^2=\frac{f_{,R}}{1+\frac{f_{,T}}{\kappa^2}}=e^{\sqrt{\frac{2}{3}}\kappa(\phi_1-\phi_2)}\,,
\end{equation}
where
\begin{eqnarray}
\kappa\phi_1\equiv\sqrt{\frac{3}{2}}\ln f_{,R} \,\,,\hspace{2.1cm} \\
\kappa\phi_2\equiv\sqrt{\frac{3}{2}}\ln
(1+\frac{f_{,T}}{\kappa^2})\,\,.\hspace{1.2cm}
\end{eqnarray}
By the conformal transformations and using the relation
$\sqrt{-\tilde{g}}=\Omega^4 \sqrt{-g}$, the action is rewritten as
\begin{eqnarray}
S=\int
d^4x\bigg\{\sqrt{-\tilde{g}}\bigg(\frac{\tilde{R}}{2\kappa^2}-
\frac{1}{2}\tilde{g}^{\mu\nu}{\nabla_\mu
\phi_1}{\nabla_\nu \phi_1}-\frac{1}{2}\tilde{g}^{\mu\nu}{\nabla_\mu
\phi_2}{\nabla_\nu \phi_2}+\tilde{g}^{\mu\nu}{\nabla_\mu
\phi_1}{\nabla_\nu \phi_2}\hspace{1cm} \nonumber\\
-V(\phi_1,\phi_2)\bigg)-2\tilde{\lambda}
\bigg({\sqrt{-\tilde{g}}}{e^{-2\sqrt{2/3}\kappa(\phi_1
		-\phi_2)}-1}\bigg)\bigg\}+\int d^4x
\sqrt{-\tilde{g}}{\cal{L}}_{m}(\Omega^{-2}\tilde{g}_{\mu\nu},\psi_m)\,.
\end{eqnarray}
where $\tilde{\lambda}=\frac{\lambda}{2\kappa^2}$ and $V(\phi_1,\phi_2)=\frac{\Omega^2R-f}{2\kappa^2\Omega^4}$ is the potential of the scalar fields. The action (121) shows that the matter Lagrangian is coupled with conformal factor $\Omega^2$. Varying the action (121) with respect to the Lagrange multiplier we obtain $\sqrt{-\tilde{g}}=e^{2\sqrt{2/3}\kappa(\phi_1 -\phi_2)}$ which is not constant contrary to the Jordan frame. The equation of motions for the scalar fields $\phi_1$ and $\phi_2$ become~\cite{Raj17}
\begin{eqnarray}
\tilde{{\Box}} \phi_1-\tilde{\Box}
\phi_2-V_{,\phi_1}+4\tilde{\lambda}\sqrt{\frac{2}{3}}\kappa
e^{-2\sqrt{\frac{2}{3}}\kappa(\phi_1-\phi_2)}-\frac{1}{\sqrt{6}}\kappa
\tilde{T}^{m}=0\,,
\end{eqnarray}
\begin{eqnarray}
\tilde{\Box} \phi_2-\tilde{\Box}
\phi_1-V_{,\phi_2}-4\tilde{\lambda}\sqrt{\frac{2}{3}}\kappa
e^{-2\sqrt{\frac{2}{3}}\kappa(\phi_1-\phi_2)}+\frac{1}{\sqrt{6}}\kappa
\tilde{T}^{m}=0\,.
\end{eqnarray}
We obtain the field equations by variation of the action (121) with respect to the metric $\tilde{g}_{\mu\nu}$
\begin{eqnarray}
\tilde{R}_{\mu\nu}-\frac{1}{2}\tilde{g}_{\mu\nu}\tilde{R}=
\kappa^2(\tilde{T}_{\mu\nu}+\tilde{T}^{m}_{\mu\nu})\,,
\end{eqnarray}
where $\tilde{T}_{\mu\nu}$ is the energy-momentum of the scalar fields
\begin{eqnarray}
\tilde{T}_{\mu\nu}(\phi_1,\phi_2)
=\nabla_{\mu}\phi_1\nabla_{\nu}\phi_1+\nabla_{\mu}\phi_2\nabla_{\nu}\phi_2
-2\nabla_{\mu}\phi_1\nabla_{\nu}\phi_2-
\tilde{g}_{\mu\nu}\Big[\frac{1}{2}\tilde{g}^{\alpha\beta}\nabla_{\alpha}
\phi_1\nabla_{\beta}\phi_1+\hspace{0.6cm}\nonumber \\
\frac{1}{2}\tilde{g}^{\alpha\beta}\nabla_{\alpha}\phi_2\nabla_{\beta}\phi_2-
\tilde{g}^{\alpha\beta}\nabla_{\alpha}\phi_1\nabla_{\beta}\phi_2+V(\phi_1,\phi_2)\Big]
-2\tilde{\lambda}\tilde{g}_{\mu\nu}e^{-2\sqrt{2/3}\kappa(\phi_1
	-\phi_2)}\,.
\end{eqnarray}
As we have shown in Ref.~\cite{Raj17}, the Lagrange multiplier in the Einstein frame is a constant ($\tilde{\lambda}=\tilde{\lambda}_0$), while in Jordan frame this quantity is not constant and varies with time. Now, the effective potential is defined as
\begin{equation}
V_{eff}(\phi_1,\phi_2)=\frac{{\Omega^2}R-f}{2\kappa^2{\Omega^4}}+2\tilde{\lambda}_0
e^{-2\sqrt{\frac{2}{3}}\kappa(\phi_1-\phi_2)}\,.
\end{equation}
The FRW metric (4) takes the following form in the Einstein frame
\begin{eqnarray}
d\tilde{s}^2=\Omega^2
ds^2=-\tilde{a}^{-6}(\tilde{\tau})d\tilde{\tau}^2+\tilde{a}^2(\tilde{\tau})dx_idx^i\,.
\end{eqnarray}
By conformal transformation, the scale factor, time variable and energy-momentum tensor of matter in the Einstein frame transform respectively as
\begin{eqnarray}
\tilde{a}=e^{\frac{1}{\sqrt{6}}\,\kappa (\phi_1-\phi_2)}\,a\quad\quad \tilde{T}_{\mu\nu}^m=e^{-\sqrt{\frac{2}{3}}\kappa (\phi_1-\phi_2)}\,T_{\mu\nu}^m\,,\quad\quad d\tilde{\tau}=e^{2\sqrt{\frac{2}{3}}\kappa(\phi_1-\phi_2)}\,d\tau\,.
\end{eqnarray}
Thus, the field equations in the Einstein frame reduce to
\begin{eqnarray}
\tilde{{\cal{H}}}^2=\frac{\kappa^2}{3}\bigg[\frac{1}{2}\dot{\phi}_{1}^{2}
+\frac{1}{2}\dot{\phi}_{2}^{2}-\dot{\phi}_{1}\dot{\phi}_{2}+\tilde{a}^{-6}
V_{eff}(\phi_1,\phi_2)+\tilde{a}^{-6}\tilde{\rho}_m\bigg]\,\,,
\end{eqnarray}
\begin{eqnarray}
-2\dot{\tilde{{\cal{H}}}}-9\tilde{{\cal{H}}}^2=\kappa^2\bigg[\frac{1}{2}\dot{\phi}_{1}^{2}
+\frac{1}{2}\dot{\phi}_{2}^{2}-\dot{\phi}_{1}\dot{\phi}_{2}-\tilde{a}^{-6}
V_{eff}(\phi_1,\phi_2)+\tilde{a}^{-6}\tilde{p}_m\bigg]\,\,.
\end{eqnarray}
where
\begin{eqnarray}
\tilde{\cal{H}}=e^{-2\sqrt{\frac{2}{3}}\kappa(\phi_1-\phi_2)}\bigg[{\cal{H}}+\frac{1}{\sqrt{6}}\kappa(\frac{d\phi_1}{d\tau}-\frac{d\phi_2}{d\tau})\bigg]\,.
\end{eqnarray}
The equation of motion for the scalar fields get
\begin{eqnarray}
(\ddot{\phi}_1-\ddot{\phi}_2)\tilde{a}^6-6\tilde{{\cal{H}}}(\dot{\phi}_1-\dot{\phi}_2)\tilde{a}^6+V_{,\phi_1}^{eff}=\frac{1}{\sqrt{6}}\kappa(\tilde{\rho}_m-3\tilde{p}_m)\,,
\end{eqnarray}
\begin{eqnarray}
-(\ddot{\phi}_2-\ddot{\phi}_1)\tilde{a}^6+6\tilde{{\cal{H}}}(\dot{\phi}_2-\dot{\phi}_1)\tilde{a}^6-V_{,\phi_2}^{eff}=\frac{1}{\sqrt{6}}\kappa(\tilde{\rho}_m-3\tilde{p}_m)\,.
\end{eqnarray}
From the Jordan frame's viewpoint, there is no inflaton field in unimodular $f(R,T)$ gravity. Also, finding the solutions of the cosmic expansion in the Jordan frame is so complicate. Therefore, we use the solution of a scalar-field motion in the Einstein frame and investigate the dynamics of the inflation and reheating in this case. We consider the Starobinsky $f(R,T)$ gravity as $f(R,T)=R+\alpha R^2+2\gamma T$~\cite{Mor16} which is capable to describe the radiation era of the universe. In this regard, the Ricci scalar becomes as
\begin{eqnarray}
R=\frac{1}{2\alpha}\Big(e^{\sqrt{\frac{2}{3}}\kappa \phi_1 }-1\Big)\,,
\end{eqnarray}
and the conformal parameter is
\begin{eqnarray}
\Omega^2=F_0 \,e^{\sqrt{\frac{2}{3}}\kappa \phi_1}\,,
\end{eqnarray}
where $F_0=(1+\frac{2\gamma}{\kappa^2})^{-1}$ and
\begin{eqnarray}
\kappa \phi_1=\sqrt{\frac{3}{2}}\ln(1+2\alpha R)\,,\quad\quad\kappa \phi_2=\sqrt{\frac{3}{2}} \ln(1+\frac{2\gamma}{\kappa^2})=constant \,.
\end{eqnarray}
If we choose $\lambda_0 =\frac{1}{8\alpha F_0^3}\frac{(F_0-1)^2}{2F_0-1}$ and by considering the radiation as the ordinary matter ($p=\frac{\rho}{3}$), the effective potential is obtained as
\begin{eqnarray}
V_{eff}(\phi_1)=\frac{2F_0-1}{8\kappa \alpha F_0^2}\bigg[1-\frac{F_0}{2F_0-1}e^{-\sqrt{\frac{2}{3}}\kappa \phi_1}\bigg]^2\,.
\end{eqnarray}

\subsection{Slow roll approximation}
The inflaton rolls into the potential at $\tilde{t} = \tilde{t}_i$ ($t=t_i$) with small kinetic energy compared to the
height of the potential. By slow roll approximation the field equations and motion equations are given as
\begin{eqnarray}
\tilde{\cal{H}}=\kappa \tilde{a}^{-3}\sqrt{\frac{V_{eff}}{3}}\,,\hspace{2.5cm}
\end{eqnarray}
\begin{eqnarray}
\dot{\tilde{{\cal{H}}}}=-\kappa^2 \tilde{a}^{-6} V_{eff}\,,\hspace{2.5cm}
\end{eqnarray}
\begin{eqnarray}
\dot{\phi}_1=-\frac{(V_{eff})_{,\phi_1}\,\tilde{a}^{-3}}{2\kappa \sqrt{3V_{eff}}}\,,\hspace{2.2cm}
\end{eqnarray}
where a dot denotes derivative with respect to the time variable $\tilde{\tau}$ in Einstein frame. We assume the initial condition as $\phi_1=\phi_{1i}$. By inserting the effective potential into equation (140) and by using the relation $d\tilde{\tau}=\tilde{a}^3 d\tilde{t}$ we find
\begin{eqnarray}
\phi_1(\tilde{t})=\sqrt{\frac{3}{2}}\frac{1}{\kappa}\ln\Big[h_{i}^2-\frac{2(\tilde{t}-\tilde{t}_{i})}{3\sqrt{6\alpha(2F_0-1)}}\Big]\,,\hspace{1.6cm}
\end{eqnarray}
\begin{eqnarray}
\frac{d\phi_1}{d\tilde{t}}=-\frac{1}{3\kappa \sqrt{\alpha(2F_0-1)}}\Big[h_{i}^2-\frac{2(\tilde{t}-\tilde{t}_{i})}{3\sqrt{6\alpha(2F_0-1)}}\Big]^{-1}\,,
\end{eqnarray}
where we have set $h_{i}={e^{\frac{\kappa \phi_{1i}}{\sqrt{6}}}}$. Also, we can obtain the Hubble parameter and the scale factor in the Einstein frame as follows
\begin{eqnarray}
\tilde{H}=\frac{1}{2F_0}\sqrt{\frac{2F_0-1}{6\alpha}}\bigg[1-\frac{F_0}{2F_0-1}\Big(h_{i}^2-\frac{2(\tilde{t}-\tilde{t}_{i})}{3\sqrt{6\alpha(2F_0-1)}}\Big)^{-1}\bigg]\,,
\end{eqnarray}
\begin{eqnarray}
\tilde{a}=\tilde{a}_{i}e^{\frac{1}{2F_0}\sqrt{\frac{2F_0-1}{6\alpha}}(\tilde{t}-\tilde{t}_{i})}\Big(h_{i}^2-\frac{2(\tilde{t}-\tilde{t}_{i})}{3\sqrt{6\alpha(2F_0-1)}}\Big)^{\frac{3}{4}}\,.
\end{eqnarray}
Now, by using equation (128) and integrating, we derive the expression for time in the Jordan frame $t$ in terms of $\tilde{t}$ in order to translate the above equations in the language of the Jordan frame
\begin{eqnarray}
t=t_{i}-3\sqrt{\frac{6\alpha(2F_0-1)}{F_0}}\bigg[\Big(h_{i}^2-\frac{2(\tilde{t}-\tilde{t}_{i})}{3\sqrt{6\alpha(2F_0-1)}}\Big)^{\frac{1}{2}}-h_{i}\bigg]\,,
\end{eqnarray}
and the time in the Einstein frame gets
\begin{eqnarray}
\tilde{t}=\tilde{t}_{i}+2\sqrt{F_0}(t-t_{i})\bigg[h_{i}-\frac{1}{6}\sqrt{\frac{F_0}{6\alpha(2F_0-1)}}(t-t_i)\bigg]\,.
\end{eqnarray}
So, by substituting the above equation in the solutions (141)-(144) we find the analytic solutions in terms of the Jordan frame quantities.

The energy density of the
created particles for $f(R,T)=R+\alpha R^2+\gamma T$ is given as
\begin{eqnarray}
\rho_r(t)=\frac{N \epsilon }{1152 \pi a^4}\int_{-\infty}^{t} dt' a^4 R^2\,.
\end{eqnarray}
where $\epsilon=\frac{1}{\sqrt{24\alpha}}$ and $N$ is the number of fields that can be excited by the cosmological oscillation. By inserting $R$ from equation (134) and using equations (141), (144) and (146) we obtain
\begin{eqnarray}
\rho_r(t)=\frac{N \epsilon}{1152\pi}\hspace{11cm}\nonumber\\
\times \frac{3\sqrt{6\alpha}}{16\big[s^2-\frac{3(2F_0-1)}{F_0}h_i^2\big]}\bigg\{\sqrt{\frac{1}{3}s^2-\frac{2F_0-1}{F_0}h_i^2}\Big[A_1s^4+A_2s^2+A_3h_i^4+A_4h_i^2+A_5\Big]\nonumber\\
+\sqrt{3\pi}e^{s^2-\frac{3(2F_0-1)}{F_0}h_i^2}A_6\,\text{erf}\bigg[\sqrt{s^2-\frac{3(2F_0-1)}{F_0}h_i^2}\bigg]\bigg\}\quad\quad
\end{eqnarray}
where $s\equiv \sqrt{\frac{2(2F_0-1)}{3F_0}}\Big[\sqrt{\frac{F_0}{6\alpha(2F_0-1)}}(t-t_i)-3h_i\Big]$ and $\text{erf}(x)$ is the error function. Also we have
\begin{eqnarray}
A_1=\frac{2F_0}{9\alpha^2(2F_0-1)^2}\,,\hspace{6.2cm}\nonumber\\
A_2=\frac{8}{3\alpha^2(2F_0-1)}h_i^2-\frac{8(h_i^2-1)}{\sqrt{6\alpha}\alpha(2F_0-1)}-\frac{5F_0}{9\alpha^2(2F_0-1)^2}\,,\nonumber\\
A_3=-\frac{2}{\alpha^2 F_0}\,,\,\hspace{7.45cm}\nonumber\\
A_4=\frac{5}{3\alpha^2(2F_0-1)}-\frac{4}{\alpha^2F_0}\,,\,\hspace{5cm}\nonumber\\
A_5=-\frac{2(F_0-1)}{\alpha^2F_0(2F_0-1)}-\frac{5F_0}{6\alpha^2(2F_0-1)^2}\,,\,\hspace{2.95cm}\nonumber\\
A_6=\frac{29F_0^2-36F_0+12}{36\alpha^2F_0(2F_0-1)^2}\,.\,\hspace{5.66cm}
\end{eqnarray}
Now, we consider the boundary conditions and assume the time $\tilde{t}=\tilde{t}_0$ or $t=t_0$ when the inflation reaches $\phi\simeq 0$ for the first time. We note that the slow-roll approximation is not valid anymore in this region and the boundary conditiond are important due to the sudden transition of the potential at $\phi\simeq 0$. So, by using the analytic solution (141) we obtain
\begin{eqnarray}
\tilde{t}_0=\tilde{t}_i+\frac{3\sqrt{6\alpha(2F_0-1)}}{2}(h_i^2-1)\,,
\end{eqnarray}
and the time in the Jordan frame is
\begin{eqnarray}
t_0=t_i+3\sqrt{\frac{6\alpha(2F_0-1)}{F_0}}(h_i-1)\,.
\end{eqnarray}
Then, by using equations (142) and (144) the boundary conditions are given as
\begin{eqnarray}
\phi_{1,0}=0\,,\quad\quad \frac{d{\phi_{1,0}}}{d\tilde{t}}=-\frac{1}{3\kappa\sqrt{\alpha(2F_0-1)}}\,, \quad \quad \tilde{a}_0=\tilde{a}_ie^{\frac{3(2F_0-1)}{4F_0}(h_i^2-1)}\,.
\end{eqnarray}
Finally, we obtain the energy density $\rho$ at $t=t_0$ by taking $s=-\sqrt{\frac{3(2F_0-1)}{F_0}}$ in equation (148) and assuming large $\phi_i$ limit, i.e., $h_i\gg1$
\begin{eqnarray}
\rho_{r0}=\frac{C N\epsilon}{1152\pi}\,,
\end{eqnarray}
where
\begin{eqnarray}
C=\frac{1}{32}\frac{F_0}{2F_0-1}\bigg[-\sqrt{\frac{2F_0-1}{F_0}}\frac{F_0^2+2F_0}{6\alpha^2F_0(2F_0-1)^2}+\sqrt{3\pi}A_6e^{\frac{3(2F_0-1)}{F_0}}erf(\sqrt{\frac{3(2F_0-1)}{F_0}})\bigg].
\end{eqnarray}\\

\subsection{Fast roll approximation}
Since the inflaton rolls toward the minimum of the potential, its kinetic energy is much larger than the potential energy. Hence, we use the fast-roll approximation in field equations and  motion equations
\begin{eqnarray}
\tilde{\cal{H}}=\frac{1}{6}\kappa^2\dot{\phi}_1^2\,,\quad
\end{eqnarray}
\begin{eqnarray}
\dot{\tilde{{\cal{H}}}}=-\kappa^2\dot{\phi}_1^2\,,\quad
\end{eqnarray}
\begin{eqnarray}
\ddot{\phi}_1-6{\cal{H}}\dot{\phi}_1=0\,.
\end{eqnarray}
The inflaton falls down on bottom of the potential and oscillates in the left and right direction of the potential's minimum repeatedly. We consider the time interval $\tilde{\tau}_0<\tilde{\tau}<\tilde{\tau}_1$ for the motion of the inflaton in the left direction, while in the time interval $\tilde{\tau}_1<\tilde{\tau}<\tilde{\tau}_2$ the inflaton moves to the right direction of the potential's minimum. Then we obtain the analytic solution in each domain. At first we consider interval $\tilde{\tau}_0<\tilde{\tau}<\tilde{\tau}_1$ and by eliminating $\dot{\phi}_1$ from equations (155) and (156) we derive the Hubble parameter and the scale factor as
\begin{eqnarray}
\tilde{\cal{H}}(\tilde{\tau})=\frac{\tilde{\cal{H}}_0}{6{\tilde{\cal{H}}_0}(\tilde{\tau}-\tilde{\tau}_0)+1}\,,
\end{eqnarray}
\begin{eqnarray}
\tilde{a}(\tilde{\tau})=\tilde{a}_0\big[6{\tilde{\cal{H}}_0}(\tilde{\tau}-\tilde{\tau}_0)+1\big]^{1/6}\,.
\end{eqnarray}
By using the scale factor (159) and relation $d\tilde{t}=\tilde{a}^{-3}d\tilde{\tau}$ we obtain
\begin{eqnarray}
\tilde{t}-\tilde{t}_0=\frac{1}{3{\tilde{\cal{H}}}_0\tilde{a}_0^3}\big[6{\tilde{\cal{H}}_0}(\tilde{\tau}-\tilde{\tau}_0)+1\big]^{1/2}\,,
\end{eqnarray}
and the inverse function is
\begin{eqnarray}
\tilde{\tau}-\tilde{\tau}_0=\frac{9{\tilde{\cal{H}}}_0^2\,\tilde{a}_0^6\,(\tilde{t}-\tilde{t}_0)^2-1}{6{\tilde{\cal{H}}}_0}\,.
\end{eqnarray}
From equation (155) we have $\dot{\phi}_1=\pm \sqrt{6}\kappa \tilde{\cal{H}}$ which $\pm$ corresponds to the left and right direction of the inflaton motion respectively. For $\tilde{\tau}_0<\tilde{\tau}<\tilde{\tau}_1$, we consider $\dot{\phi}_1=- \sqrt{6}\kappa \tilde{\cal{H}}$. Hence, we get
\begin{eqnarray}
\phi_1(\tilde{\tau})=-\frac{1}{\sqrt{6}\kappa}\ln\big[6{\tilde{\cal{H}}_0}(\tilde{\tau}-\tilde{\tau}_0)+1\big]\,.
\end{eqnarray}
Then, by taking $\dot{\phi}_1=+\sqrt{6}\kappa \tilde{\cal{H}}$ for interval $\tilde{\tau}_1<\tilde{\tau}<\tilde{\tau}_2$, the solutions are given by
\begin{eqnarray}
\tilde{\cal{H}}(\tilde{\tau})=\frac{\tilde{\cal{H}}_1}{6{\tilde{\cal{H}}_1}(\tilde{\tau}-\tilde{\tau}_1)+1}\,,\hspace{3.5cm}
\end{eqnarray}
\begin{eqnarray}
\tilde{a}(\tilde{\tau})=\tilde{a}_1\big[6{\tilde{\cal{H}}_1}(\tilde{\tau}-\tilde{\tau}_1)+1\big]^{1/6}\,,\hspace{2.7cm}
\end{eqnarray}
\begin{eqnarray}
\phi_1(\tilde{\tau})=\frac{1}{\sqrt{6}\kappa}\ln\big[6{\tilde{\cal{H}}_1}(\tilde{\tau}-\tilde{\tau}_1)+1\big]-\sqrt{\frac{3}{2}}\frac{1}{\kappa}\ln{b}\,.
\end{eqnarray}
By matching the conditions between the two intervals we get $\phi_1(\tilde{\tau}_1)=-\sqrt{\frac{3}{2}}\frac{1}{\kappa}\ln{b}$, and by substituting this into equation (162) we derive
\begin{eqnarray}
\tilde{\tau}_1=\tilde{\tau}_0+\frac{b^3-1}{6\tilde{\cal{H}}_0}\,,
\end{eqnarray}
\begin{eqnarray}
\tilde{\cal{H}}_1=\tilde{\cal{H}}_0\,b^{-3}\,,
\end{eqnarray}
\begin{eqnarray}
\tilde{a}_1=\tilde{a}_0\,b^{1/2}\,.
\end{eqnarray}
By inserting these equations into the solutions (163)-(165), we obtain the same expressions for
$\tilde{\cal{H}}(\tilde{\tau})$ and $\tilde{a}(\tilde{\tau})$ as in equations (158) and (159), but $\phi_1(\tilde{\tau})$ is different from equation (162). It is given by
\begin{eqnarray}
\phi_1(\tilde{\tau})=\frac{1}{\sqrt{6}\kappa}\ln\big[6{\tilde{\cal{H}}_0}(\tilde{\tau}-\tilde{\tau}_0)+1\big]-2\sqrt{\frac{3}{2}}\frac{1}{\kappa}\ln{b}\,.
\end{eqnarray}
Then by repeating this procedure, we get the following boundary conditions for general $n$
\begin{eqnarray}
\tilde{\tau}_n=\tilde{\tau}_0+\frac{b^{3n}-1}{3\tilde{\cal{H}}_0}\,,\hspace{3.6cm}
\end{eqnarray}
\begin{eqnarray}
\tilde{\cal{H}}_n=\tilde{\cal{H}}_0 b^{-3n}\,,\hspace{4.2cm}
\end{eqnarray}
\begin{eqnarray}
\tilde{a}_n=\tilde{a}_0b^{\frac{1}{2}n}\,,\hspace{4.5cm}
\end{eqnarray}
\begin{eqnarray}
\phi_n(\tilde{\tau})=\left\{\begin{array}{ll}0\, \hspace{2.5cm} n=even\\ \\
-\sqrt{\frac{3}{2}}\frac{1}{\kappa}\ln b\hspace{1cm} n=odd\,.\\
\end{array}\right.
\end{eqnarray}
Hence, general solutions for time interval $\tilde{\tau}_{n-1}<\tilde{\tau}<\tilde{\tau}_n$ become
\begin{eqnarray}
\tilde{\cal{H}}(\tilde{\tau})=\frac{\tilde{\cal{H}}_0}{6{\tilde{\cal{H}}_0}(\tilde{\tau}-\tilde{\tau}_0)+1}\,,\hspace{8.2cm}
\end{eqnarray}
\begin{eqnarray}
\tilde{a}(\tilde{\tau})=\tilde{a}_0\big[6{\tilde{\cal{H}}_0}(\tilde{\tau}-\tilde{\tau}_0)+1\big]^{1/6}\,,\hspace{7cm}
\end{eqnarray}
\begin{eqnarray}
\phi_1(\tilde{\tau})=\left\{\begin{array}{ll}
-\frac{1}{\sqrt{6}\kappa}\ln\big[6{\tilde{\cal{H}}_0}(\tilde{\tau}-\tilde{\tau}_0)+1\big]+(n-1)\sqrt{\frac{3}{2}}\frac{1}{\kappa}\ln{b}\hspace{1cm} n=odd\\ \\
\frac{1}{\sqrt{6}\kappa}\ln\big[6{\tilde{\cal{H}}_0}(\tilde{\tau}-\tilde{\tau}_0)+1\big]-n\sqrt{\frac{3}{2}}\frac{1}{\kappa}\ln{b}\hspace{2.2cm} n=even
\end{array}\right.
\end{eqnarray}
\begin{eqnarray}
\dot{\phi}_1(\tilde{\tau})=\left\{\begin{array}{ll}
-\sqrt{6}\kappa^{-1}\tilde{\cal{H}}(\tilde{\tau})\hspace{0.7cm} n=odd\\ \\
\sqrt{6}\kappa^{-1}\tilde{\cal{H}}(\tilde{\tau})
\hspace{1cm} n=even
\end{array}\right.\hspace{5.3cm}
\end{eqnarray}
These equations show that the Hubble parameter and scale factor are independent of the direction of inflaton motion in oscillation around the potential's minimum while $\phi$ and $\dot{\phi}$ have different solutions for the different parity of $n$. By using equation (161) we can rewrite the equation (174)-(176)
in the following forms
\begin{eqnarray}
\tilde{H}(\tilde{t})=\frac{1}{3(\tilde{t}-\tilde{t}_0)}\,,\hspace{9.5cm}
\end{eqnarray}
\begin{eqnarray}
\tilde{a}(\tilde{t})=\tilde{a}_0\big[3{\tilde{\cal{H}}_0}(\tilde{t}-\tilde{t}_0)\big]^{1/3}\,,\hspace{8cm}
\end{eqnarray}
\begin{eqnarray}
\phi_1(\tilde{t})=\left\{\begin{array}{ll}
-\frac{2}{\sqrt{6}\kappa}\ln\big[3{\tilde{\cal{H}}_0}(\tilde{t}-\tilde{t}_0)\big]+(n-1)\sqrt{\frac{3}{2}}\frac{1}{\kappa}\ln{b}\hspace{1cm} n=odd\\ \\
\frac{2}{\sqrt{6}\kappa}\ln\big[3{\tilde{\cal{H}}_0}(\tilde{t}-\tilde{t}_0)\big]-n\sqrt{\frac{3}{2}}\frac{1}{\kappa}\ln{b}\hspace{2.2cm} n=even\,.
\end{array}\right.\hspace{1.2cm}
\end{eqnarray}
Now, we can write the Einstein frame time $\tilde{t}$ in terms of Jordan frame time $t$ by integrating $e^{-\frac{1}{\sqrt{6}}\kappa \phi_1}$ and convert the above solutions to those in the Jordan frame. Jordan frame time is obtained
\begin{eqnarray}
t(\tilde{t})=\left\{\begin{array}{ll}
t_{n-1}+\frac{b^{-\frac{n-1}{2}}}{4{\cal{H}}_0\sqrt{F_0}}\Big\{\big[3{\tilde{\cal{H}}_0}(\tilde{t}-\tilde{t}_0)\big]^{4/3}-b^{2(n-1)}\Big\}\hspace{1.3cm} n=odd\\ \\
t_{n-1}+\frac{b^{\frac{n}{2}}}{2{\cal{H}}_0\sqrt{F_0}}\Big\{\big[3{\tilde{\cal{H}}_0}(\tilde{t}-\tilde{t}_0)\big]^{2/3}-b^{n-1}\Big\}\hspace{1.7cm} n=even\,.
\end{array}\right.
\end{eqnarray}
Using equation (128), (131), (178), (179) and (181) we obtain the Hubble parameter and scale factor in the Jordan frame respectively as
\begin{eqnarray}
a(t)=\left\{\begin{array}{ll}
\frac{\tilde{a}_0 b^{\frac{n-1}{2}}}{\sqrt{F_0}}\big[4\sqrt{F_0}\tilde{\cal{H}}_0b^{-\frac{3}{2}(n-1)}(t-t_{n-1})+1\big]^{1/2} \hspace{1.5cm} n=odd\\ \\
\frac{\tilde{a}_0 b^{\frac{n}{2}}}{\sqrt{F_0}} \hspace{7.7cm} n=even
\end{array}\right.\hspace{1.8cm}
\end{eqnarray}

\begin{eqnarray}
H(t)=\left\{\begin{array}{ll}
\frac{3}{2}{\sqrt{F_0}}\tilde{\cal{H}}_0 b^{-\frac{3}{2}(n-1)}\big[4\sqrt{F_0}\tilde{\cal{H}}_0b^{-\frac{3}{2}(n-1)}(t-t_{n-1})+1\big]^{-1} \hspace{1.5cm} n=odd\\ \\
\frac{1}{2}{\sqrt{F_0}}\tilde{\cal{H}}_0 b^{-(n-1)}\big[4\sqrt{F_0}\tilde{\cal{H}}_0b^{-\frac{3}{2}(n-1)}(t-t_{n-1})+1\big]^{-1} \hspace{1.7cm} n=even\,.
\end{array}\right.
\end{eqnarray}
From equation (170) and using $d\tilde{t}=a^{-3}d\tilde{\tau}$ we have $\ln({\tilde{t}-\tilde{t}_0})\propto n$. Note that since the inflaton moves to the right and left direction during oscillation, the time intervals are equal in terms of the Einstein frame's time.  Using equation (181) we get $\ln{(t-t_0)\simeq\zeta \ln{(\tilde{t}-\tilde{t}_0)}}$ where $\zeta=\frac{4}{3}$ and  $\zeta=\frac{2}{3}$ are for the left and right direction of inflaton motion respectively. Therefore, the left directed region is twice as long as the right directed region in terms of time duration of Jordan frame. So, we can find the average power of the Jordan frame quantities. For example, the average power for time duration is $(4/3 + 2/3)/(1 + 1) = 1$, as time duration in Einstein frame is proportional to time duration in Jordan frame on average, i.e., $\langle t-t_0\rangle\propto \tilde{t}-\tilde{t}_0$. For scale factor we have $\ln{a}=\eta \ln{t}$ where $\eta=1/2$ and $0$ are for left and right direction regions respectively. Thus, the average power is $(1/2\times 2+0\times 1)/)(2+1)=1/3$ which we can write $\langle a(t)\rangle\propto (t-t_0)^{1/3}$. Hence, we obtain the averaged Hubble parameter as
\begin{eqnarray}
\langle H(t)\rangle=\frac{1}{3(t-t_0)}\,.
\end{eqnarray}
Average scale factor is given by integrating from equation (160)
\begin{eqnarray}
\langle a(t)\rangle=a_0[H_{0}(t-t_0)]^{1/3}\,.
\end{eqnarray}
In the slow roll region, the scale factor is quasi-de Sitter whereas the scale factor in the fast roll region is as $a(t)\propto (t-t_0)^{1/2}$ and $a(t)\simeq const$. The fast-roll approximation does not hold at late time and the inflaton loses its kinetic energy. So, the inflaton reaches a false vacuum at $\phi_1=0$.

Now by using the analytic solutions, we estimate the reheating temperature as a function of the parameters of the model. The oscillation of scalaron ends quickly before radiation dominates the universe. At first, we compare the energy density of radiation due to particle creation and the energy density of gravity. So we get approximately the energy density of radiation, $\rho_r$, as
\begin{eqnarray}
\langle \rho_r(t)\rangle=\rho_{{r_{0}}}\bigg(\frac{\langle a(t)\rangle}{a_0}\bigg)^{-4}\approx \frac{\rho_{{r_{0}}}}{[3H_{0}(t-t_0)]^{4/3}}\,.
\end{eqnarray}
By substituting the function $f(R,T)=R+\alpha R^2+\gamma T$ into equation (14), we can write the equation of motion in the Jordan frame as follows
\begin{eqnarray}
3H^2=\kappa^2\Big[\big(1+\frac{4\gamma}{3\kappa^2}\big)\rho_r+\rho_g\Big]\,,
\end{eqnarray}
where $\rho_g$ is the effective energy density of gravity in Jordan frame which is defined as
\begin{eqnarray}
\rho_g=\frac{1}{\kappa^2}\Big[18\alpha(\dot{H}+2H^2)(\dot{H}+2H^2)+3H+\lambda\Big]\,,
\end{eqnarray}
where $\rho_r$ is negligible compared to $\rho_g$. Then the energy density of gravity becomes
\begin{eqnarray}
\langle \rho_g(t)\rangle\approx 3\langle H\rangle^2\approx\frac{1}{3(t-t_0)^2}\,,
\end{eqnarray}
where we have used equation (184). At the end of inflation, the radiation is dominated in the universe and its energy decays as $\rho_r\propto a^{-4}$. With the scale factor evolving as $a\propto t^{1/2}$, the energy density of radiation evolves as $\rho_r\propto t^{-2}$.

To estimate the time at which reheating terminates, that is, $t_{reh}$, and the reheating temperature, $T_{reh}$, we set $\langle \rho_r\rangle=\langle \rho_g \rangle$. Using equation (186) and (189) we derive
\begin{eqnarray}
t_{reh}-t_0\approx \frac{\sqrt{3}H_{0}^{2}}{(\rho_{r0})^{3/2}}\approx \sqrt{3}H_{0}^{2}\Big(\frac{CN\epsilon}{1152\pi}\Big)^{-3/2}\,.
\end{eqnarray}
Next, we can relate the temperature at the reheating era to the scale factor in this era by using the following equation
\begin{eqnarray}
\frac{T_{reh}}{T_0}=\bigg(\frac{43}{11g_{reh}}\bigg)^{1/3}\bigg(\frac{a_0}{\langle a_{reh}\rangle}\bigg)\approx \bigg(\frac{43}{11g_{reh}}\bigg)^{1/3}\frac{\sqrt{\rho_{r0}}}{\sqrt{3}H_{0}}\approx\bigg(\frac{43}{11g_{reh}}\bigg)^{1/3}(\sqrt{3}H_{0})^{-1}\Big(\frac{C N\epsilon}{1152\pi}\Big)^{1/2}\,.
\end{eqnarray}
These equations depend on $\rho_{{r_{0}}}$, so the end of the reheating is sensitive to the boundary conditions at the
transition from the slow roll regime to the fast roll regime.\\

To have a numerical estimation in this framework, with $g_{reh}=106.75$, $\alpha=-0.2$ and $\gamma=0.8$, we determine the reheating time from Eq. (190) as $t_{reh}\approx 1.9 \times 10^{-29} s$ and reheating temperature from (191) as $T_{reh}\approx 1.5 \times 10^{11}$ GeV. These are well in the range obtained in other models as have been mentioned previously. 

\section{Summary and Conclusion}
Following our previous construction of the Unimodular $f(R,T)$ gravity~\cite{Raj17} and the Energy Conditions in Unimodular $f(R,T)$ gravity~\cite{Raj21}, in this paper we reconsidered  cosmological inflation and the issue of reheating in the unimodular $f(R,T)$ gravity. At the first step, we analyzed in details the embedding of cosmological inflation within unimodular $f(R, T)$ gravity. In confrontation with observational data from Planck Satellite, this model is observationally viable in some subspaces of the model parameter space. Then we studied the issue of reheating after the cosmic inflation. The rapid oscillation phase after the slow-roll regime is studied in details and the numbers of e-folds during radiation and recombination eras are calculated analytically in addition to some numerical estimations of the main physical quantities. By gathering all obtained analytical results, we derived the reheating temperature in terms of $T_{CMB}$, scalar spectral index and the power spectrum. We have also studied the issue of particle creation in this model which is affected by the non-conservation of energy-momentum in this generalized framework. We have shown that massless particles can be produced during radiation dominated phase in this scenario. Then we have investigated the inflation and reheating dynamics in Einstein frame for unimodular Starobinsky $f(R,T)$ model. The analytical solutions in slow-roll inflation and fast-roll oscillation regimes have been derived in this scenario following by some numerical estimations. In Summary:

\begin{itemize}
\item We considered a power-law potential $V(\phi)=q \phi^{n}$ for the scalar field which has led to an exponential scale factor, showing the possibility of positively accelerated expansion of the universe in this model. As we have shown, by adopting suitable initial conditions, it is possible to realize intermediate inflation in this scenario.

\item We have analyzed the behavior of inflation for the simplest unimodular $f(R, T)$ gravity of the form $f(R, T) =R+\beta T$ with $\lambda={\alpha_1}R+{\alpha_2}T$, coupled with a scalar field. We found that the inflation observables are independent of model parameters $\beta$, $\alpha_{1}$ and $\alpha_{2}$, but are very sensitive to the number of e-folds $N$ and the parameter $n$. Such parameters obtained in our setup are compatible with the latest set of data from Planck Collaboration.

\item To determine the reheating temperature, we divided the evolution of the universe into different segments (sub-eras) and obtained the corresponding e-folds number for each segment and then summed over all segments. We obtained an analytical expression for the reheating temperature in terms of the CMB temperature, the spectral index, the power spectrum and the parameters of the model, Eq. (69). With $T_{CMB}=2.725$ K$ =2.348\times10^{-4}$ eV, $g_{reh}=106.75$, $k_0=0.002M pc^{-1}$ (with $Mpc^{-1}=6.39\times 10^{-30}eV$) and by using  WMAP7 results as ${\cal{P}}_s(k_0)=2.441^{+0.088}_{-0.092}\times 10^{-9}$ and $n_s=0.963\pm 0.012$, we obtained $T_{reh}=2.125\times 10^{12}$ GeV which is larger than the result obtained in the standard general relativity framework.

\item We investigated the problem of massless particle creation in unimodular $f(R, T)$ homogeneous and isotropic cosmologies for an exact power law solution for the scale factor of the universe and imposing viable initial conditions. We have calculated the number of created particles for some values of $m$ where $a(t)\propto t^{m}$. It has been proved that for each $k$, there is a critical value of the conformal time $\eta$ for which the number density of the created particles grows abruptly. We noticed that for each $k$ there is an abrupt growth of the number of created particles at the initial time and after a brief oscillation, it turns constant and remains so throughout the cosmic evolution. For smaller values of $k$ the amount of created particles is growing, that is, a huge amount of particles were created with low $k$ in the past. This means that higher $k$ modes essentially contribute much less to the total number of created particles than the small $k$ modes.

 \item We studied also thermodynamics of irreversible processes in the presence of gravitationally created particles. In this case, non-conservation of the matter stress-energy tensor $\nabla^{\mu}T_{\mu\nu}\neq 0$ plays an important role and implies a transfer of energy from the geometry to the matter sector of the theory towards gravitational particle creation on cosmological scales. To proceed, we adopted the formalism of irreversible thermodynamics of open systems in the presence of matter creation. In this framework, creation of matter gravitationally acts as a source of internal energy. Then we computed the created particle number density along with some numerical estimation. Following Ref.~\cite{Hashiba2019}, for a stable, purely gravitationally created particle as dark matter (called $X$ in the text), in order to explain sufficient reheating and also the present dark matter abundance in the same time, it is required that $m_{X}\sim10^{3}$GeV and $n_{X}\approx 2.0\times 10^{-8}$ GeV$^{3}$ at the decoupling time by the freeze-in process.

\item We derived the analytic solutions in the slow-roll inflation and the fast-roll oscillation regimes in the Einstein frame. We obtained physical quantities in the Jordan frame by the inverse conformal transformations. In the slow-roll regime, the scale factor undergoes quasi-de Sitter expansion. In the fast-roll regime, the inflaton oscillates two times inside the potential plateau. An interesting feature of this model is that the averaged time evolution of a scale factor is proportional to $t^{\frac{1}{3}}$ because of the periodic abrupt changes of the Hubble parameter.\\
    Finally, we calculated the time and temperature of the reheating stage in terms of the parameters of the model analytically with numerical estimations as $t_{reh}\approx 1.9 \times 10^{-29} s$ and $T_{reh}\approx 1.5 \times 10^{11}$ GeV.

\end{itemize}

\end{document}